\documentclass[letterpaper,12pt]{article}

\usepackage{dgjournal}          

\usepackage{mathptmx}

\usepackage{graphics}

\usepackage[authoryear,comma,longnamesfirst,sectionbib]{natbib} 
\usepackage{bm}
\usepackage{natbib}
\usepackage{latexsym}
\usepackage{color}

\usepackage{here}

\begin{document}
\baselineskip 7mm 


\begin{center}
{\Large \bf Counterfactual Reasoning with Disjunctive Knowledge in a Linear Structural Equation Model}
\vspace*{10mm}

{MANABU KUROKI}\\
{Department of Data Science, Institute of Statistical Mathematics \\
10-3 Midori-cho, Tachikawa, Tokyo 190-8562, Japan \\
{mkuroki@ism.ac.jp}}
\vspace*{5mm}

\end{center}

\begin{abstract}
{We consider the problem of estimating counterfactual quantities when prior knowledge is available in the form of disjunctive statements.
These include disjunction of conditions (e.g., ``the patient is more than 60 years of age") as well as disjuction of antecedants (e.g., ``had the patient taken either drug A or drug B"). 
Focusing on linear structural equation models (SEM) and imperfect control plans, we extend the counterfactual framework of  Balke and Pearl (1995) , Chen and Pearl (2015), and Pearl (2009, pp. 389-391) from unconditional to conditional plans, from a univariate treatment to a set of treatments, and from point type knowledge to disjunctive knowledge. 
Finally, we provide improved matrix representations of the resulting counterfactual parameters, and improved computational methods of their evaluation.}
\end{abstract}

{\noindent}Keywords: (causal) path diagram; disjunctive control plan; non-recursive structural equation model; stochastic control plan; total effect.
\vspace*{10mm}

\section{Introduction}


Counterfactual reasoning, which is widely used in practical science, has played an important role in treatment estimation, lawsuit compensation for hazardous exposure, and planning and policy analysis. 
For example, the counterfactual statement ``if I had taken aspirin, my headache would have been gone by now" implies that, in the real world, because I did not take aspirin, I still have a headache. 
In addition, this statement implicitly compares two outcomes: the actual outcome that I still have a headache because I did not take aspirin, and the counterfactual one that my headache would have gone if I had taken aspirin. 
This comparison of the actual and counterfactual outcomes enables us to evaluate the treatment effect of aspirin intake on headache recovery. 
This type of counterfactual reasoning is {often} generalized to the problem of evaluating the (unconditional) counterfactual query ``how would response variable $Y$ change, had treatment $X$ been $x$ (counterfactually)?". This has been discussed widely in the literature on causal inference (Imbens and Rubin, 2015; Morgan and Winship, 2007; Pearl, 2009; Rubin, 2006). 

In this paper, we generalize problems of unconditional counterfactual queries and consider the evaluation of the conditional counterfactual query ``how would response variable $Y$ change, if the set of treatments $\mbox{\boldmath $X$}$ were controlled by the values of other variables (counterfactually), given that we know $H$ (actually)?", where proposition $H$ denotes the knowledge derived in the real world, which we make explicit to facilitate the analysis. 
The importance of conditional counterfactual reasoning in practical science is worth emphasizing. 
For example, in the field of quality control, for defective products whose characteristic values fall outside the control limits, we often wish to know how the quality characteristics of the defective products would change if a certain quality improvement plan were carried out, before actually carrying it out (Kuroki, 2012). 
In social science, Chen and Pearl (2014) considered the situation where they wish to estimate the effect on test scores of a school policy that requires students who are lazy in doing their homework to attend the afterschool program. 
Such counterfactual reasoning emphasizes an understanding of the generic laws in practical science and shapes future decision making more than unconditional counterfactual reasoning. 

Conditional counterfactual reasoning has been studied in the context of ``probabilities of causation" by many researchers in medical science (Greenland and Robins, 1988; Robins, 2004; Robins and Greenland, 1989ab), artificial intelligence (Pearl, 1999; Tian and Pearl, 2000ab), social science (Dawid et al., 2014; Yamamoto, 2012), risk analysis (Cai and Kuroki, 2005), and statistics (Kuroki and Cai, 2011). 
Based on a wider context than ``probabilities of causation", Balke and Pearl (1994ab, 1995) presented a formal notation, semantics, and three-step computational algorithm including an Abduction step, Action step, and Prediction step, to evaluate counterfactual queries, and provided computational methods for the counterfactual distribution. 
However, to the best of our knowledge, since Balke and Pearl's (1995) study, there has been little discussion of conditional counterfactual queries based on linear structural equation models (SEMs) .
Balke and Pearl (1995) formulated counterfactual quantities based on the distributional characteristics of random disturbances, and required SEM researchers and practitioners to understand the computational algorithm to evaluate the counterfactual quantities under the assumption of Gaussian random disturbances. 
Thus, it would be difficult for SEM researchers and practitioners to apply their results to empirical studies. 

Based on this background, we consider the problem of clarifying how the mean vector and covariance matrix would change if a set of treatments $\mbox{\boldmath $X$}$ were controlled by the values of covariates, intermediate variables, and/or a response variable (counterfactually), when prior knowledge is available in the form of disjunctive knowledge of certain variables (actually). 
{These include disjunction of conditions (e.g., ``the patient is more than 60 years of age"). }
In practical science, there are many situations where we need to focus on distributional characteristics if the conditional plan of $\mbox{\boldmath $X$}$ that is, the counterfactual antecedent ``if $\mbox{\boldmath $X$}$ were controlled by the values of other variables" were carried out (Kuroki, 2012; Murphy, 2003; Pearl, 2009). 
However, most of the results of previous studies have focused on an unconditional plan given by the counterfactual antecedent ``if $\mbox{\boldmath $X$}$ was a specific constant vector $\mbox{\boldmath $x$}$". 
To achieve our aim, {under linear structural equation models (SEM) and the imperfect control plans}, we extend the counterfactual framework provided by Balke and Pearl (1995), Chen and Pearl (2014), and Pearl (2009, pp.~389-391), from {unconditional to a conditional plans}, from a univariate treatment to a set of treatments, and from {point type knowledge} to disjunctive knowledge. 
Here, the imperfect control plan, which represents that $\mbox{\boldmath $X$}$ is controlled by the values of other variables with errors, may be considered as another type of disjunctive plans discussed by Pearl (2017) such as disjuction of antecedants (e.g., ``had the patient taken either drug A or drug B"). 
Moreover, we formulate the mean vector and covariance matrix if the imperfect control plan of $\mbox{\boldmath $X$}$ is carried out (counterfactually), given that prior knowledge is available in the form of disjunctive knowledge of some variables (actually). 
Different from the work by Balke and Pearl (1995), this paper provides an explicit expression of the mean vector and covariance matrix of the counterfactual distribution without the assumption of Gaussian random disturbances. 
Moreover, these can be evaluated without executing the computational algorithms developed by Balke and Pearl (1994ab, 1995). 
Compared with Balke and Pearl's computational algorithm, such explicit expressions enable us to clarify their properties making it easy for SEM researchers and practitioners to apply their results to empirical studies. 
Additionally, in contrast to the work by Balke and Pearl (1995), when we consider a perfect control plan, neither the mean vector nor the covariance matrix of random disturbances appears in our formulations. 
Thus, our results can help SEM researchers and practitioners not only reduce the computational effort of evaluating counterfactual quantities, but also understand the causal mechanisms of how the distributional characteristics of the response variable would change if some treatments were controlled in a given subpopulation.

\section{Preliminaries}

\subsection{Linear Structural Equation Model}

A directed graph is a pair $G=(\mbox{\boldmath $V$}, \mbox{\boldmath $E$})$, where $\mbox{\boldmath $V$}=\{V_{1},\cdots,V_{n_{v}}\}$ is a finite set of vertices and the set $\mbox{\boldmath $E$}$ of arrows is a subset of the set $\mbox{\boldmath $V$}{\times}\mbox{\boldmath $V$}$ of ordered pairs of distinct vertices. 
$n_{v}$ represents the number of elements in $\mbox{\boldmath $V$}$, with a similar notation used for other numbers. 
Regarding the graph theoretic terminology used in this paper, the reader should refer to textbooks covering graphical models (Edwards, 2000; Lauritzen, 1996; Whittaker, 2009) and graphical causal models (Pearl, 2009; Spirtes et al., 2000). 

When a directed graph $G=(\mbox{\boldmath $V$}, \mbox{\boldmath $E$})$ is given with a set $\mbox{\boldmath $V$}$ of variables, graph $G$ is called a (causal) path diagram if each child-parent family in graph $G$ represents a linear SEM describing the data generating process:
\begin{equation}
V_{i}=\mu_{v_{i}{\cdot}{\rm pa}(v_{i})}+\sum_{V_{j}{\in}{\rm pa}(V_{i})}\alpha_{i j}V_{j}+\epsilon_{i},\hspace{.1cm}i=1,\ldots,n_v , 
\label{1}
\end{equation}
where ${\rm pa}(V_{i})$ denotes the set of parents of $V_{i}$ in $G$ and an exogenous random disturbance $\epsilon_{i}$ is assumed to have mean $0$ and variance $\sigma_{\epsilon_i \epsilon_i}$ $(i=1,... ,n_v)$. 
The covariance between $\epsilon_i$ and $\epsilon_{j}$ $(i\neq j; i,j=1,\ldots,n_v)$, is denoted as $\sigma_{\epsilon_i \epsilon_j}$ if it exists. 
Additionally, both $\mu_{v_{i}{\cdot}{\rm pa}(v_{i})}$ and $\alpha_{ij}$ are constant values, and $\alpha_{ij}({\neq}0)$ is called a path coefficient or the direct effect of $V_{j}$ on $V_{i}$. 
It should be noted that the set $\mbox{\boldmath $V$}$ of variables can contain both observed and unobserved variables, which are endogenous variables in the sense that they are affected by random disturbances. 
In addition, if a directed graph includes directed cycles, the corresponding SEM is said to be non-recursive; otherwise, it is said to be recursive. 
Furthermore, parameters $\mu_{v_{i}{\cdot}{\rm pa}(v_{i})}$, $\alpha_{i j}$, $\sigma_{\epsilon_i \epsilon_i}$, and $\sigma_{\epsilon_i \epsilon_j}$ $(i,j=1,...,n_v;i\neq j)$ are assumed to be independent of the values of $V_1,...,V_{n_v}, \epsilon_1,...,\epsilon_{n_v}$. 
{Under} such an assumption, the random disturbances do not need to follow the Gaussian distribution. 
For a detailed discussion of linear SEMs, refer, for example, to Bollen (1989), J$\ddot{\mbox{o}}$reskog (1979), and Pearl (2009). 

The total effect $\tau_{v_{i}v_{j}}$ of $V_{j}$ on $V_{i}$ is defined as the sum of the products of the path coefficients on the sequence of arrows along all directed paths from $V_{j}$ to $V_{i}$. 
In particular, the total effect of a set of variables $\mbox{\boldmath $X$}$ on another set of variables $\mbox{\boldmath $Y$}$, denoted as $\mbox{\boldmath $\tau$}_{yx}$, is defined as the matrix whose $(i,j)$ component is the sum of the products of the path coefficients on the sequence of arrows along all directed paths from $V_{j}{\in}\mbox{\boldmath $X$}$ to $V_{i}{\in}\mbox{\boldmath $Y$}$, but not those passing through $\mbox{\boldmath $X$}{\backslash}\{V_{j}\}$. 

\subsection{Stability Condition}

Letting $A_{vv}$ be the path coefficient matrix $A_{vv}=(\alpha_{ij})_{1\leq i,j\leq n_v}$ in Equation (\ref{1}), linear SEM (\ref{1}) can be rewritten as 
\begin{equation}
\mbox{\boldmath $V$}=\mbox{\boldmath $\mu$}_{v{\cdot}{\rm pa}(v)}+A_{vv}\mbox{\boldmath $V$}+\mbox{\boldmath $\epsilon$}_{v}, 
\label{221}
\end{equation}
where $\mbox{\boldmath $\epsilon$}_{v}=(\epsilon_{1},\cdots,\epsilon_{n_v})'$ and $\mbox{\boldmath $\mu$}_{v{\cdot}{\rm pa}(v)}=(\mu_{v_{1}{\cdot}{\rm pa}(v_1)},$
$\cdots,\mu_{v_{n_v}{\cdot}{\rm pa}(v_{n_{v}})})'$. 
Here, the transposed vector/matrix is represented by the prime notation ($'$). 
If $I_{n_v,n_v}$ represents an $n_{v}{\times}n_{v}$ identity matrix, with similar notation used for other identity matrices, there are many representations equivalent to Equation (\ref{221}). 
Letting $A^{0}_{vv}=I_{n_v,n_v}$ and substituting the right-hand side of Equation (\ref{221}) for $\mbox{\boldmath $V$}$ on the right-hand side yields 
\begin{eqnarray*}
\mbox{\boldmath $V$}&=&\mbox{\boldmath $\mu$}_{v{\cdot}{\rm pa}(v)}+A_{vv}\mbox{\boldmath $V$}+\mbox{\boldmath $\epsilon$}_{v}=(I_{n_v,n_v}+A_{vv})(\mbox{\boldmath $\mu$}_{v{\cdot}{\rm pa}(v)}+\mbox{\boldmath $\epsilon$}_{v})+A^{2}_{vv}\mbox{\boldmath $V$}, 
\end{eqnarray*}
which is true if Equation (\ref{221}) is true. 
Performing this operation $k$ times yields
\begin{displaymath}
\mbox{\boldmath $V$}=\sum^{k-1}_{i=0}A^{i}_{vv}(\mbox{\boldmath $\mu$}_{v{\cdot}{\rm pa}(v)}+\mbox{\boldmath $\epsilon$}_{v})+A^{k}_{vv}\mbox{\boldmath $V$}. 
\label{223} 
\end{displaymath}
If both $A^{k}_{vv}$ and ${\displaystyle \sum^{k-1}_{i=0}A^{i}_{vv}}$ converge to their respective matrices, the linear SEM is said to be stable (Bentler and Freeman, 1983). 
Here, matrix $A_{vv}$ is said to be convergent (Ben-Israel and Greville, 1972) if the following equation holds: 
\begin{displaymath}
\lim_{k{\rightarrow}\infty}A^{k}_{vv}=\mbox{\boldmath $0$}_{n_v,n_v}, 
\end{displaymath}
where $\mbox{\boldmath $0$}_{n_v,n_v}$ is an $n_v{\times}n_v$ zero matrix. 
Similar notation is used for other zero matrices. 
It is known that matrix $A_{vv}$ is convergent {if and only if} the maximum values of the absolute values of all eigenvalues of matrix $A_{vv}$ are less than one (Bentler and Freeman, 1983). 
In addition, Bentler and Freeman (1983) stated that 
\begin{displaymath}
(I_{n_v,n_v}-A_{vv})^{-1}=\sum^{\infty}_{k=0}A^{k}_{vv}
\end{displaymath}
is true if $A_{vv}$ is a convergent matrix. 

Stability {implies} that the observed data were generated from the equilibrium or steady state distribution of an underlying process. 
{Under} this situation, it is possible to consider carrying out the control plan described in the next section, because the mean vector and covariance matrix can be evaluated. 
Thus, this paper assumes that cause--effect relationships can be described as linear SEMs under the stability condition. 
Here, it is noted that a linear recursive SEM satisfies the stability condition. 
That is, under the stability condition, we can provide a unified discussion of the counterfactual quantities in linear recursive and non-recursive SEMs. 

\section{Counterfactual Analysis}

\subsection{Control Plan}

Consider the (possibly non-recursive) data generating process depicted in Fig.~1. 
\begin{figure}[hhh]

\hspace*{\fill}\hspace*{1cm}\resizebox{7cm}{!}{\includegraphics{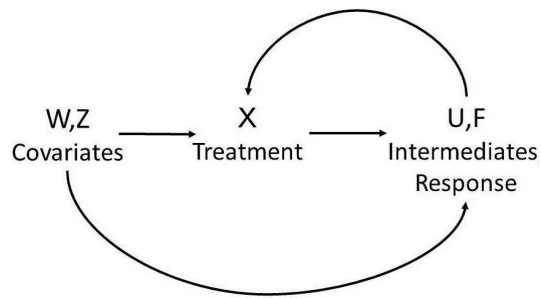}}\hspace*{\fill}

\caption{Data generating process}

\hspace*{\fill}\hspace*{\fill}
\end{figure}

In Fig.~1, $\mbox{\boldmath $X$}$ represents a set of treatments, which can be controlled by the states of the covariates, intermediate variables, and/or response variable. 
$\mbox{\boldmath $U$}\cup\mbox{\boldmath $F$}$ are sets of variables, including the intermediate variables and a response variable of interest, $Y$, which are affected by at least one element of $\mbox{\boldmath $X$}$ and may have an effect on some elements of $\mbox{\boldmath $X$}$. 
In addition, $\mbox{\boldmath $W$}$ and $\mbox{\boldmath $Z$}$ represent sets of covariates that cannot be affected by $\mbox{\boldmath $X$}$. 

If nonexperimental data are available and the cause--effect relationships between variables can be described as a linear SEM, we consider carrying out the control plan in which $\mbox{\boldmath $X$}$ is controlled by the values of the other variables, according to the following linear function:
\begin{equation}
\mbox{\boldmath $X$}=\mbox{\boldmath $x$}+\mbox{\boldmath $a$}\mbox{\boldmath $F$}+\mbox{\boldmath $b$}\mbox{\boldmath $W$}+\mbox{\boldmath $\epsilon$}^{\ast}_{x}=h(\mbox{\boldmath $F$}, \mbox{\boldmath $W$},\mbox{\boldmath $\epsilon$}^{\ast}_{x}),
\label{312}
\end{equation}
where $\mbox{\boldmath $x$}$ is a constant vector and both $\mbox{\boldmath $a$}$ and $\mbox{\boldmath $b$}$ are constant matrices that can be determined by an operator. 
If both $\mbox{\boldmath $a$}$ and $\mbox{\boldmath $b$}$ are zero matrices, the control plan is called an unconditional plan; otherwise, it is called a conditional plan. 
In addition, $\mbox{\boldmath $\epsilon$}^{\ast}_{x}$ is a random disturbance vector with mean vector $\mbox{\boldmath $0$}_{n_x, 1}$ and covariance matrix $\Sigma_{\epsilon^{\ast}_{x} \epsilon^{\ast}_{x}}$ for carrying out the control plan of $\mbox{\boldmath $X$}$ (Kuroki, 2012). 

In this paper, letting $R_h$ be a subset of {the given values} of $\mbox{\boldmath $H$}\subset \mbox{\boldmath $V$}$, we assume that $\mbox{\boldmath $\epsilon$}^{\ast}_{x}$ is independent of other random disturbances in the counterfactual world in which control plan (\ref{312}) would be carried out for the subpopulation satisfying $\mbox{\boldmath $H$}\in R_h$. 
Furthermore, both $\mbox{\boldmath $F$}$ and $\mbox{\boldmath $W$}$ are used to carry out the control plan of $\mbox{\boldmath $X$}$ (Kuroki, 2012; Kuroki and Miyakawa, 2003). 
Although some elements of $\mbox{\boldmath $U$}$ and $\mbox{\boldmath $Z$}$ may be observed to evaluate the total effect of $\mbox{\boldmath $X$}$ on $Y$ in many situations, they are not used to carry out the control plan of $\mbox{\boldmath $X$}$. 
If $\mbox{\boldmath $\epsilon$}^{\ast}_{x}$ does not exist in Equation (\ref{312}), Equation (\ref{312}) is said to be perfect; otherwise, it is imperfect. 
An imperfect control plan {implies} that some treatments could not be manipulated exactly owing to random disturbances such as physical impossibility or performance cost. 
It is important to evaluate the mean vector and covariance matrix when carrying out an imperfect control plan, because we cannot always achieve a perfect control plan. 

Based on the settings above, 
the aim of this paper is to investigate how the mean vector and covariance matrix of $\mbox{\boldmath $S$}=\mbox{\boldmath $F$}\cup \mbox{\boldmath $U$}$ would change if control plan (\ref{312}) were carried out (counterfactually), given that we know $\mbox{\boldmath $H$}\in R_h$ in the real world, i.e., the mean vector, $E(\mbox{\boldmath $S$}|{\rm do}(\mbox{\boldmath $X$}=h(\mbox{\boldmath $F$},\mbox{\boldmath $W$},\mbox{\boldmath $\epsilon$}^{\ast}_{x})),\mbox{\boldmath $H$}\in R_h)$, and the covariance matrix, var$(\mbox{\boldmath $S$}|{\rm do}(\mbox{\boldmath $X$}=h(\mbox{\boldmath $F$},\mbox{\boldmath $W$},\mbox{\boldmath $\epsilon$}^{\ast}_{x})),\mbox{\boldmath $H$}\in R_h)$. 
Here, ${\rm do}(\mbox{\boldmath $X$}=h(\mbox{\boldmath $F$},\mbox{\boldmath $W$},\mbox{\boldmath $\epsilon$}^{\ast}_{x}))$ denotes that the equations for $\mbox{\boldmath $X$}$ in Equation (2) are set to $\mbox{\boldmath $X$}=h(\mbox{\boldmath $F$},\mbox{\boldmath $W$},\mbox{\boldmath $\epsilon$}^{\ast}_{x})$ through external intervention (Pearl, 2009). 

\subsection{Total Effects}

It {would be} worthwhile stating the relationship between the definitions of total effect in linear SEMs and causal effect defined as $E[Y|\mbox{do}(X = x)]$ in nonlinear ones. 

Letting $X$ be a univariate treatment, $E[Y|\mbox{do}(X = x)]$ is interpreted as the expected value of $Y$ when setting the equations for $\mbox{\boldmath $X$}$ in Equation (2) to $\mbox{\boldmath $X$}=\mbox{\boldmath $x$}$ through external intervention (Pearl, 2009). 
Mathematically, such an intervention is represented by removing the structural equation for $X$ and replacing it with equality $X = x$. 
Then, the relationships between $X$ and its prior causes are deleted by this intervention, and statistical dependence between $X$ and $Y$ is generated by the directed paths from $X$ to $Y$. 
In particular, in linear SEMs, such a dependence is generated by the sum of the products of the path coefficients on the sequence of arrows along all directed paths from $X$ to $Y$, which is equivalent to the total effect stated in Section 2.1. 
Here, when $E[Y|\mbox{do}(X = x)]$ is differentiable relative to some reference point $x$, in nonlinear SEMs, the average causal effect of $X$ on $Y$ around $X=x$ $d E[Y|\mbox{do}(X = x)]/dx$ generally depends on the choice of reference point $x$. 
In contrast, in linear SEMs, it is characterized by the total effect $\tau_{yx}$ of $X$ on $Y$, which does not depend on the choice of reference point $x$. 

In the framework of structural causal models (Pearl, 2009), $E[Y|\mbox{do}(X = x)]$ and $E(Y_{x})$ are also connected through external intervention, where $Y_{x}$ represents the counterfactual sentence ``$Y$ would be $y$ had $X$ been $x$". 
The key to interpreting counterfactual sentences is to treat the subjunctive phrase ``had $X$ been $x$" as an instruction to make a ``minimal" modification to the current model, thereby ensuring the antecedent condition $X = x$. 
This minimal modification amounts to removing the structural equation for $X$ and replacing it with equality $X = x$, which is the same mathematical operation as discussed above. 
Thus, the expected value of $Y_{x}$, $E[Y_{x}]$, is given as $E[Y_{x}]=E[Y|\mbox{do}(X=x)]$. 
If $E[Y_x]$ is differentiable relative to some reference point $x$ in linear SEMs, 
$E[Y_x]$ and $\tau_{yx}$ are connected via $E[Y|\mbox{do}(X = x)]$; $d E[Y_x]/dx$ is also characterized by the total effect $\tau_{yx}$ of $X$ on $Y$.

{Here, ``do" operation induces a sequential data generating process that the intermediate variables and the response variable are observed as consequences of the external intervention. 
Thus, when intermediate variables and/or the response variable are included in $\mbox{\boldmath $H$}$, ``do" expression may be inappropriate but ``the counterfactuals" $Y_{x}$ would be better to be used for the problem setting where $\mbox{\boldmath $H$}$ are observed before the intervention (Pearl, 2009,pp.392-393). 
However, this paper uses ``do" expression to emphasize that such an intervention is represented by removing the structural equation for $X$ and replacing it with equality $X = x$ in the counterfactual world. 
}
\subsection{Procedure}

Chen and Pearl (2014) pointed out that linear SEMs can be used to answer counterfactual queries such as ``given that we observe $\mbox{\boldmath $T$} = \mbox{\boldmath $t$}$ for a given individual, what would we expect the value of $Y$ to be for that individual if $X$ were $x$?". In addition, they presented a gentle introduction to the computational algorithm for counterfactual queries given by Balke and Pearl (1995). 

Considering Chen and Pearl's observation, to formulate the mean vector $E(\mbox{\boldmath $S$}|{\rm do}(\mbox{\boldmath $X$}=h(\mbox{\boldmath $F$},\mbox{\boldmath $W$},\mbox{\boldmath $\epsilon$}^{\ast}_{x})),$
$\mbox{\boldmath $H$}\in R_h)$, and covariance matrix var$(\mbox{\boldmath $S$}|{\rm do}(\mbox{\boldmath $X$}=h(\mbox{\boldmath $F$},\mbox{\boldmath $W$},\mbox{\boldmath $\epsilon$}^{\ast}_{x})),$
$\mbox{\boldmath $H$}\in R_h)$, we extend Balke and Pearl's counterfactual framework from the case where an unconditional plan is carried out with {point type knowledge} (i.e., $\mbox{\boldmath $H$}=\mbox{\boldmath $h$}$) to that where control plan (\ref{312}) is carried out with disjunctive knowledge (i.e., $\mbox{\boldmath $H$}\in R_h$). 
The computational algorithm comprises the following three steps:

\begin{list}{}{
\renewcommand{\makelabel}{\hfill}
\setlength{\itemindent}{-3mm}
\setlength{\topsep}{0mm}
\setlength{\parsep}{0mm}
\setlength{\leftmargin}{6mm}
}
\item Step 1 (Abduction): Update the distribution of random disturbances pr$(\mbox{\boldmath $\epsilon$}_{v})$ using the evidence $\mbox{\boldmath $H$}\in R_h$, to obtain pr$(\mbox{\boldmath $\epsilon$}_{v}|\mbox{\boldmath $H$}\in R_h)$. 

\item Step 2 (Action): Modify the original SEM (\ref{221}) by replacing the structural equations for the variables in $\mbox{\boldmath $X$}$ by Equation (3).

\item Step 3 (Prediction): Use the updated probabilities of random disturbances in Step 1 together with the modified SEM in Step 2 to compute the mean vector and covariance matrix of $\mbox{\boldmath $S$}$.
\end{list}
\begin{figure}[H]

\hspace*{\fill}\resizebox{5cm}{!}{\includegraphics{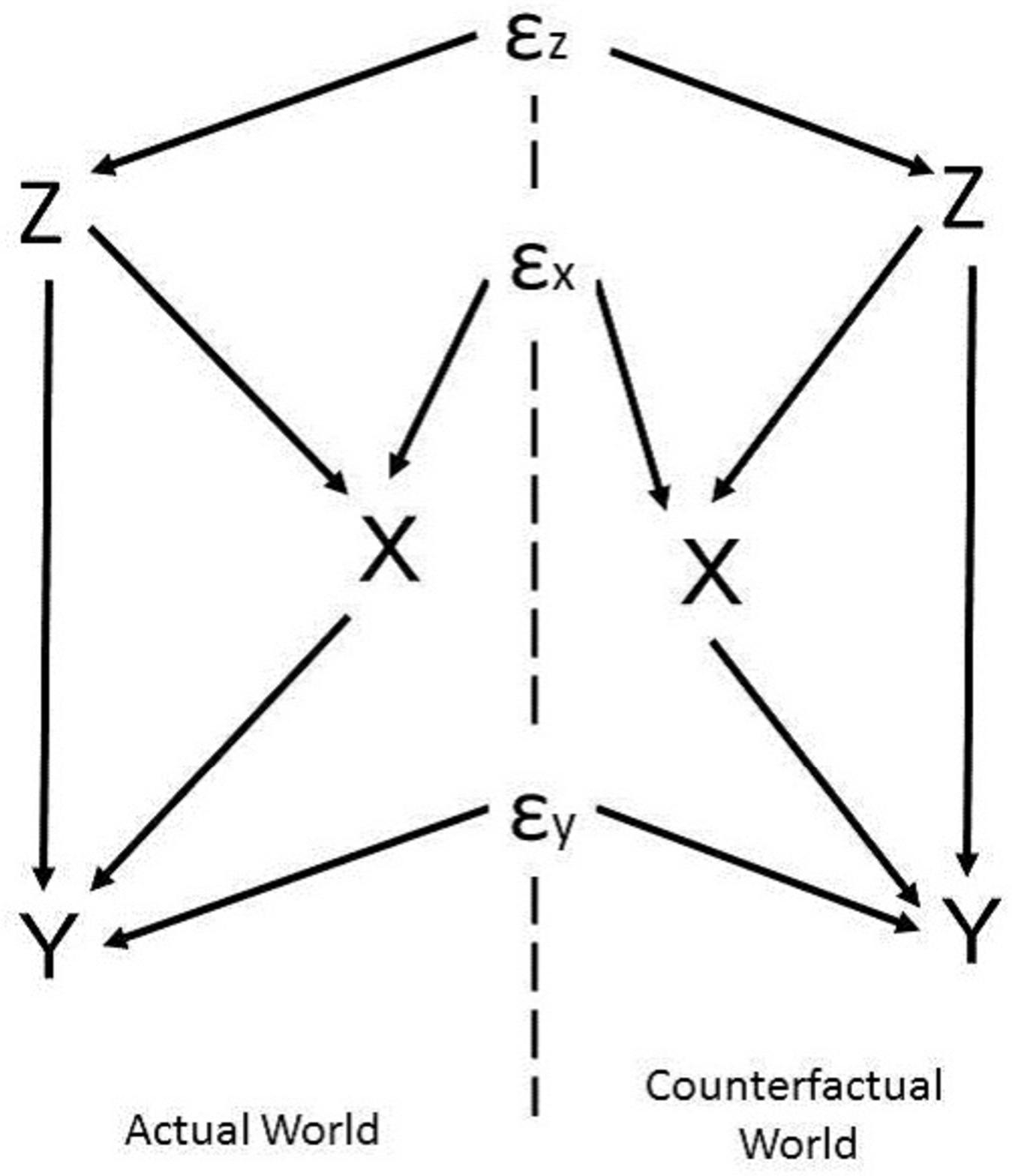}}\hspace*{\fill}
\hspace*{\fill}\resizebox{5.5cm}{!}{\includegraphics{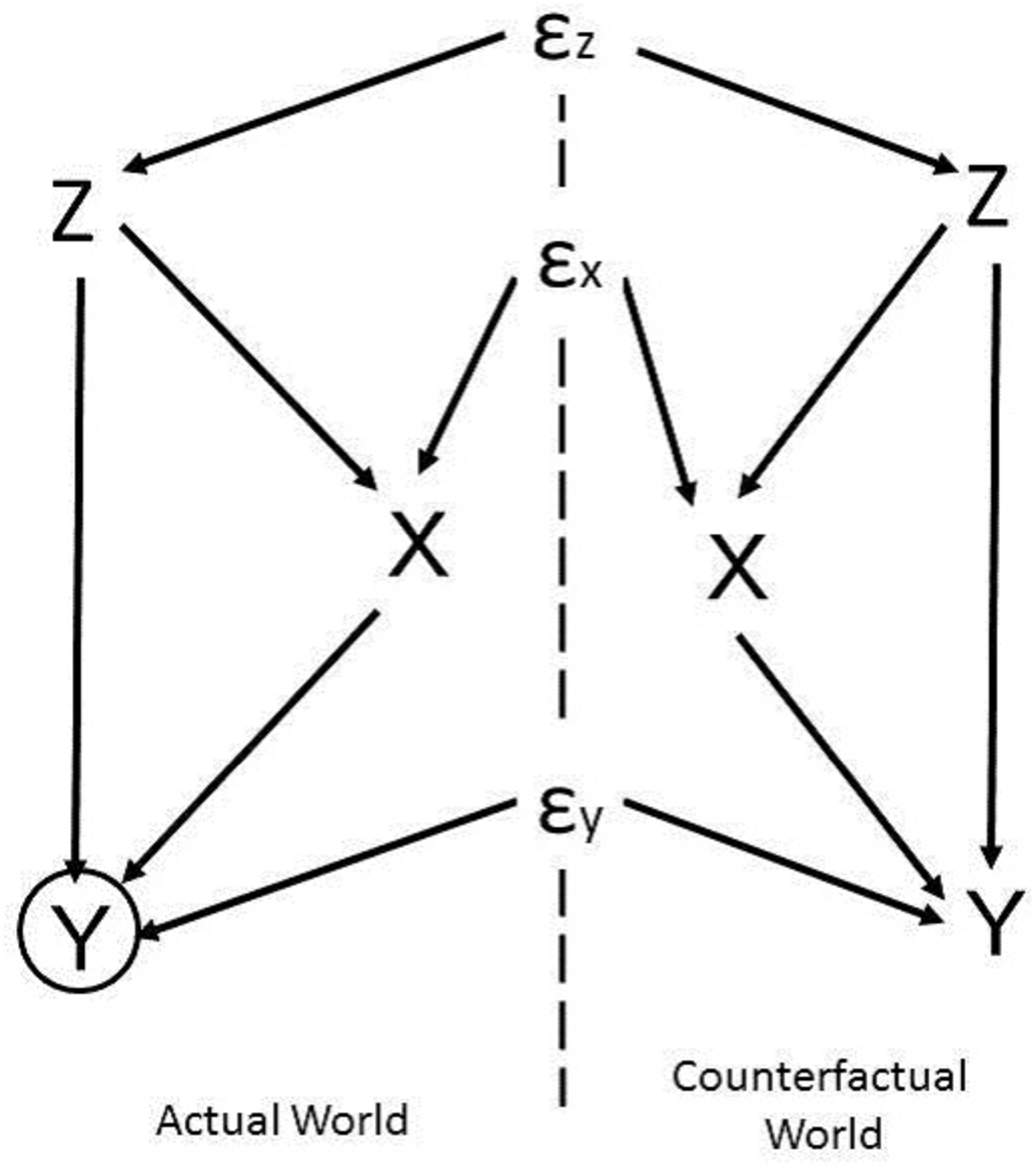}}\hspace*{\fill}
\vspace*{3mm}

\hspace*{\fill}\hspace*{\fill}(a) Setup \hspace*{\fill}\hspace*{\fill}(b) Abduction \hspace*{\fill}
\vspace*{3mm}

\hspace*{\fill}\resizebox{5.7cm}{!}{\includegraphics{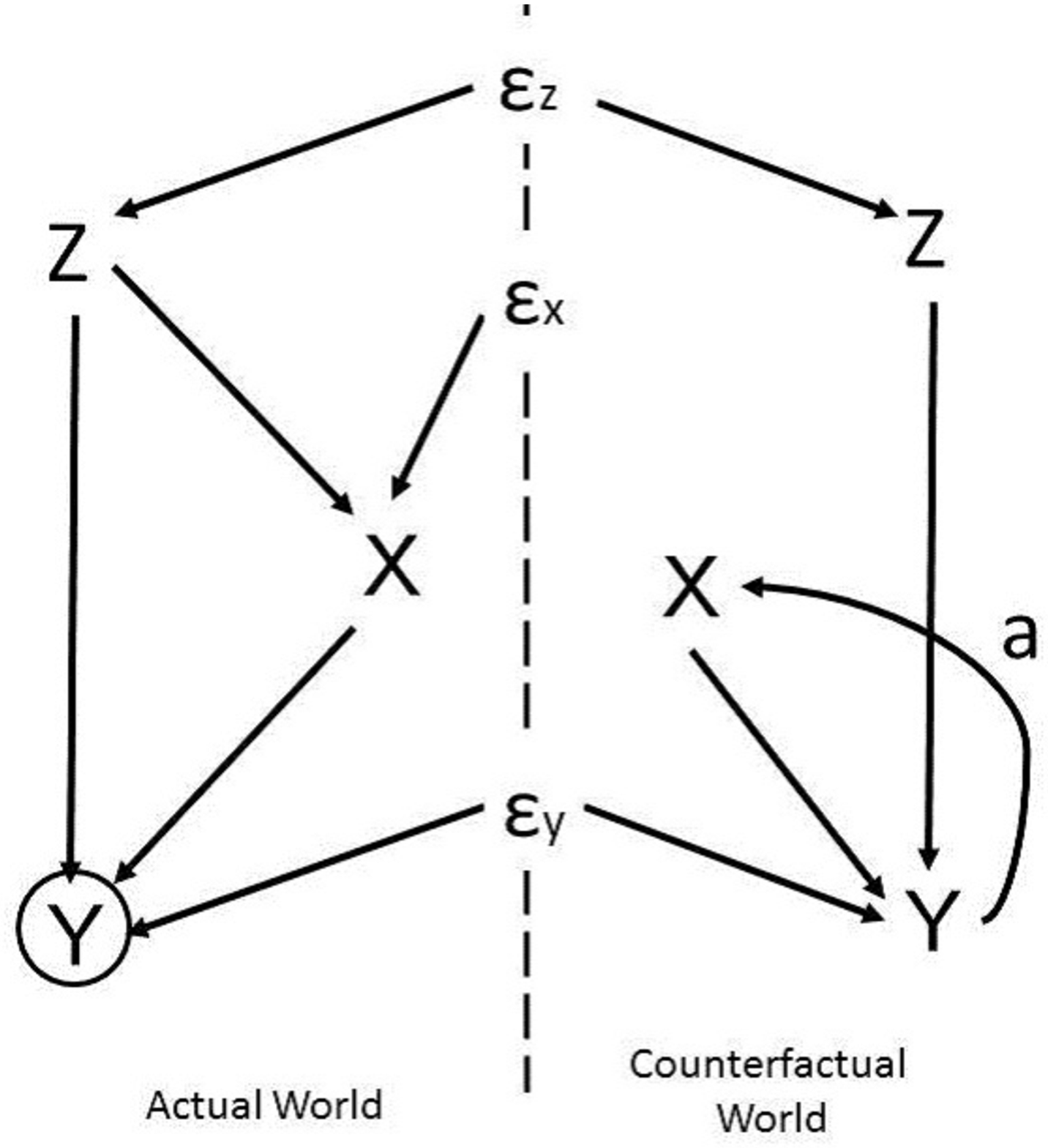}}\hspace*{\fill}
\vspace*{3mm}

\hspace*{\fill}(c) Action\hspace*{\fill}
\vspace*{3mm}

\caption{Procedure for evaluating counterfactual quantities: $\textcircled{Y}$ indicates that $Y$ is observed in the real world and is used to update the distribution of random disturbances. 
{$a$ is a coefficient of the control plan $X=a\times Y$ which is conducted in the counterfactual world. } } 
\end{figure}

These three steps are explained by using probability propagation based on the dual networks shown in Fig.~2(a): one representing the real world, and the other the counterfactual world (Balke and Pearl, 1994ab, 1995). 
In the dual network, the network structure of the real world is the same as that of the counterfactual one. 
The dual network is constructed by connecting the real world to the counterfactual world through random disturbances that are shared between both worlds.  
Then, the graphical representation of Step 2 is given by deleting the arrows pointing to $\mbox{\boldmath $X$}$ from the counterfactual world and adding arrows based on Equation (3) to the counterfactual world, as shown in Fig.~2(c). 
It should be noted that the idea of a dual network is not limited to linear functions, but applies whenever we are willing to assume the functional form of structural equations. 
For details, refer to Balke and Pearl (1994ab, 1995). 

It should also be noted that Balke and Pearl (1994ab, 1995) assumed Gaussian random disturbances to formulate the mean vector and covariance matrix of counterfactual queries. 
On the contrary, Pearl (2009, pp.~389--391) formulated the mean of counterfactual queries without the distributional assumption of random disturbances. 
In fact, the assumption of Gaussian random disturbances is not necessary to satisfy the condition that the parameters of Equation (1) are independent of the values for $V_1,...,V_{n_v}, \epsilon_1,...,\epsilon_{n_v}$, because we do not use the ``if and only if" relationship between the statistical independence and the null correlation, and we do not focus on the statistical inference problem of counterfactual queries. 
The assumption of Gaussian random disturbances satisfies the condition, but the distributions of random disturbances satisfying the requirement are not limited to the Gaussian distribution; we can assume other distributions of random disturbances if necessary. 

\section{Formulation}

\subsection{Setup}

According to the discussion in the previous section, we partition the set $\mbox{\boldmath $V$}$ of vertices in path diagram $G$ into the following three disjoint sets: 

{\noindent}$\mbox{\boldmath $S$}=\mbox{\boldmath $F$}{\cup}\mbox{\boldmath $U$}$: a set of vertices that are descendants of at least one element in $\mbox{\boldmath $X$}$ ($\mbox{\boldmath $F$}{\cap}\mbox{\boldmath $U$}=\phi$). 
Here, $\mbox{\boldmath $F$}$ and $\mbox{\boldmath $U$}$ include the first $n_{f}$ components and the next $n_{u}$ components of $\mbox{\boldmath $S$}$, respectively. In addition, a response variable of interest, $Y$, is included in either $\mbox{\boldmath $F$}$ or $\mbox{\boldmath $U$}$. 

{\noindent}$\mbox{\boldmath $X$}$: a set of treatments. 

{\noindent}$\mbox{\boldmath $T$}=\mbox{\boldmath $W$}{\cup}$\mbox{\boldmath $Z$}$=\mbox{\boldmath $V$}{\backslash}(\mbox{\boldmath $X$}{\cup}\mbox{\boldmath $S$}):$ a set of nondescendants of $\mbox{\boldmath $X$}$ ($\mbox{\boldmath $W$}{\cap}$\mbox{\boldmath $Z$}$=\phi$). 
Here, $\mbox{\boldmath $W$}$ and $\mbox{\boldmath $Z$}$ include the first $n_{w}$ components and the next $n_{z}$ components of $\mbox{\boldmath $T$}$, respectively. 

According to the above partitioning of $\mbox{\boldmath $V$}$, 
let $A_{st}$ be a path coefficient matrix of $\mbox{\boldmath $T$}$ on $\mbox{\boldmath $S$}$ whose $(i,j)$ component is the path coefficient of $T_{j}$ on $S_{i}$ ($ S_{i}\in\mbox{\boldmath $S$},T_{j}\in\mbox{\boldmath $T$}$). 
Then, Equation (\ref{221}) can be rewritten as:
\begin{eqnarray}
\left(
\begin{array}{c}
\mbox{\boldmath $S$} \\
\mbox{\boldmath $X$} \\
\mbox{\boldmath $T$}
\end{array}
\right)
&=&
\left(
\begin{array}{c}
\mbox{\boldmath $\mu$}_{s{\cdot}{\rm pa}(s)} \\
\mbox{\boldmath $\mu$}_{x{\cdot}{\rm pa}(x)} \\
\mbox{\boldmath $\mu$}_{t{\cdot}{\rm pa}(t)}
\end{array}
\right)+
\left(
\begin{array}{ccc}
A_{ss}& A_{sx}& A_{st}\\
A_{xs}& A_{xx}& A_{xt}\\
\mbox{\boldmath $0$}_{n_t,n_s}& \mbox{\boldmath $0$}_{n_t,n_x}& A_{tt}
\end{array}
\right)
\left(
\begin{array}{c}
\mbox{\boldmath $S$} \\
\mbox{\boldmath $X$} \\
\mbox{\boldmath $T$}
\end{array}
\right)+
\left(
\begin{array}{c}
\mbox{\boldmath $\epsilon$}_{s} \\
\mbox{\boldmath $\epsilon$}_{x} \\
\mbox{\boldmath $\epsilon$}_{t}
\end{array}
\right),
\label{a3}
\end{eqnarray}
where $\mbox{\boldmath $\epsilon$}_{s}$, $\mbox{\boldmath $\epsilon$}_{x}$, and $\mbox{\boldmath $\epsilon$}_{t}$ are random disturbance vectors corresponding to $\mbox{\boldmath $S$}$, $\mbox{\boldmath $X$}$, and $\mbox{\boldmath $T$}$, respectively. 
In addition, $\mbox{\boldmath $\mu$}_{s{\cdot}{\rm pa}(s)}$, $\mbox{\boldmath $\mu$}_{x{\cdot}{\rm pa}(x)}$, and $\mbox{\boldmath $\mu$}_{t{\cdot}{\rm pa}(t)}$ are constant vectors corresponding to $\mbox{\boldmath $S$}$, $\mbox{\boldmath $X$}$, and $\mbox{\boldmath $T$}$, respectively. 
Similar notation is used for other vectors. 
Then, letting
\begin{displaymath}
A_{vv}=\left(
\begin{array}{ccc}
A_{ss}& A_{sx}& A_{st}\\
A_{xs}& A_{xx}& A_{xt}\\
\mbox{\boldmath $0$}_{n_t,n_s}& \mbox{\boldmath $0$}_{n_t,n_x}& A_{tt}
\end{array}
\right), 
A_{xs,xs}=\left(
\begin{array}{cc}
A_{ss}&A_{sx}\\
A_{xs}&A_{xx}
\end{array}
\right),
\end{displaymath} 
the characteristic equation of $A_{vv}$ is given by
\begin{eqnarray*}
\lefteqn{\det(\lambda I_{n_v,n_v}-A_{vv})
=\det\left(
\begin{array}{ccc}
\lambda I_{n_s,n_s}-A_{ss}& -A_{sx}& -A_{st}\\
-A_{xs}& \lambda I_{n_x,n_x}-A_{xx}& -A_{xt}\\
\mbox{\boldmath $0$}_{n_t,n_s}& \mbox{\boldmath $0$}_{n_t,n_x}& \lambda I_{n_t,n_t}-A_{tt}
\end{array}
\right)
}\\
&=&
\det\left(
\begin{array}{cc}
\lambda I_{n_s,n_s}-A_{ss}& -A_{sx}\\
-A_{xs}& \lambda I_{n_x,n_x}-A_{xx}
\end{array}
\right)
\det(\lambda I_{n_t,n_t}-A_{tt})\\
&=&\det(\lambda I_{n_t,n_t}-A_{tt})\det(\lambda I_{n_{s}+n_{x},n_{s}+n_{x}}-A_{xs,xs})=0
\end{eqnarray*}
from the basic formula for the determinant of the block matrices. 
Thus, both $A_{tt}$ and $A_{xs,xs}$ must be convergent matrices for Equation (\ref{a3}) to satisfy the stability condition. 

\subsection{Step 1: Abduction}

Before continuing this discussion, we need to define some notation. 
For variables $X$ and $Y$ and sets of variables $\mbox{\boldmath $S$}$, $\mbox{\boldmath $W$}$, and $\mbox{\boldmath $Z$}$, let $\mu_{y}=E(Y)$, $\mbox{\boldmath $\mu$}_{w}=E(\mbox{\boldmath $W$})$, $\sigma_{xx}=$var$(X)$, $\sigma_{xy}=$cov$(X,Y)$, $\Sigma_{ww}=$var$(\mbox{\boldmath $W$})$, and $\Sigma_{sw}=$cov$(\mbox{\boldmath $S$},\mbox{\boldmath $W$})$. 
Similar notation is used for other matrices and parameters. 
First, we update both the mean vector and covariance matrix of the random disturbances through $\mbox{\boldmath $H$}\in R_{h}$. 
Generally, given linear SEMs, and because we have 
$$E(\mbox{\boldmath $V$}|\mbox{\boldmath $H$}\in R_{h})=\mbox{\boldmath $\mu$}_{v{\cdot}{\rm pa}(v)}+A  E(\mbox{\boldmath $V$}|\mbox{\boldmath $H$}\in R_{h})+E(\mbox{\boldmath $\epsilon$}_v|\mbox{\boldmath $H$}\in R_{h}),$$
from Equation (\ref{221}), we have  
\begin{eqnarray}
\lefteqn{\mbox{\boldmath $\mu$}_{\epsilon_v. r_h}=E(\mbox{\boldmath $\epsilon$}_v|\mbox{\boldmath $H$}\in R_{h})=-\mbox{\boldmath $\mu$}_{v{\cdot}{\rm pa}(v)}+(I_{n_v, n_v}-A)  E(\mbox{\boldmath $V$}|\mbox{\boldmath $H$}\in R_{h})}\nonumber \\
&=&-(I_{n_v, n_v}-A)  \mbox{\boldmath $\mu$}_{v}+(I_{n_v, n_v}-A)  \mbox{\boldmath $\mu$}_{v.r_{h}}=(I_{n_v, n_v}-A)(\mbox{\boldmath $\mu$}_{v.r_{h}}-\mbox{\boldmath $\mu$}_{v}),
\label{313}
\end{eqnarray}
noting that $\mbox{\boldmath $\mu$}_{v}=\mbox{\boldmath $\mu$}_{v{\cdot}{\rm pa}(v)}+A \mbox{\boldmath $\mu$}_{v}$ from Equation (\ref{221}), where $\mbox{\boldmath $\mu$}_{v.r_{h}}=E(\mbox{\boldmath $V$}|\mbox{\boldmath $H$}\in R_{h})$. 
In addition, since we have 
\begin{eqnarray}
\lefteqn{\mbox{var}(\mbox{\boldmath $V$}|\mbox{\boldmath $H$}\in R_{h})=A \mbox{var}(\mbox{\boldmath $V$}|\mbox{\boldmath $H$}\in R_{h})  A'} \nonumber \\
&&+A \mbox{cov}(\mbox{\boldmath $V$},\mbox{\boldmath $\epsilon$}_v|\mbox{\boldmath $H$}\in R_{h})
+\mbox{cov}(\mbox{\boldmath $\epsilon$}_v,\mbox{\boldmath $V$}|\mbox{\boldmath $H$}\in R_{h})  A'+\mbox{var}(\mbox{\boldmath $\epsilon$}_v|\mbox{\boldmath $H$}\in R_{h})
\label{311}
\end{eqnarray}
and
\begin{equation}
\mbox{var}(\mbox{\boldmath $V$}|\mbox{\boldmath $H$}\in R_{h})=\mbox{var}(\mbox{\boldmath $V$}|\mbox{\boldmath $H$}\in R_{h})  A'+\mbox{cov}(\mbox{\boldmath $V$},\mbox{\boldmath $\epsilon$}_v|\mbox{\boldmath $H$}\in R_{h}),
\label{332}
\end{equation}
by substituting Equation (\ref{312}) for Equation (\ref{311}), we have
\begin{eqnarray*}
\mbox{var}(\mbox{\boldmath $\epsilon$}_v|\mbox{\boldmath $H$}\in R_{h})&=&(I-A)\mbox{var}(\mbox{\boldmath $V$}|\mbox{\boldmath $H$}\in R_{h})(I-A)'=(I-A)\Sigma_{vv.r_{h}}(I-A)',
\end{eqnarray*}
where $\Sigma_{vv.r_{h}}=\mbox{var}(\mbox{\boldmath $V$}|\mbox{\boldmath $H$}\in R_{h})$. 

\subsection{Step 2: Action}

Next, to formulate the modified SEM through Equation (\ref{a3}) with the updated distribution of random disturbances from Step 1, we let $C_{xs}=(\mbox{\boldmath $a$};\mbox{\boldmath $0$}_{n_x,n_u})$ and $C_{xt}=(\mbox{\boldmath $b$};\mbox{\boldmath $0$}_{n_x,n_z})$. 
When carrying out the control plan, the modified linear SEM can be represented by
\begin{equation}
\left(
\begin{array}{c}
\mbox{\boldmath $S$} \\
\mbox{\boldmath $X$} \\
\mbox{\boldmath $T$}
\end{array}
\right)
=
\left(
\begin{array}{c}
\mbox{\boldmath $\mu$}_{s{\cdot}{\rm pa}(s)} \\
\mbox{\boldmath $x$}\\
\mbox{\boldmath $\mu$}_{t{\cdot}{\rm pa}(t)}
\end{array}
\right)+
\left(
\begin{array}{ccc}
A_{ss}& A_{sx}& A_{st}\\
C_{xs}& \mbox{\boldmath $0$}_{n_x,n_x}& C_{xt}\\
\mbox{\boldmath $0$}_{n_t,n_s}& \mbox{\boldmath $0$}_{n_t,n_x}& A_{tt}
\end{array}
\right)
\left(
\begin{array}{c}
\mbox{\boldmath $S$} \\
\mbox{\boldmath $X$} \\
\mbox{\boldmath $T$}
\end{array}
\right)+
\left(
\begin{array}{c}
\mbox{\boldmath $\epsilon$}_{s.r_{h}} \\
\mbox{\boldmath $\epsilon$}^{\ast}_{x} \\
\mbox{\boldmath $\epsilon$}_{t.r_{h}}
\end{array}
\right).
\label{4}
\end{equation}
Noting that $\mbox{\boldmath $\mu$}_{s{\cdot}{\rm pa}(s)}=(I_{n_s,n_s}-A_{ss})\mbox{\boldmath $\mu$}_{s}-A_{sx}\mbox{\boldmath $\mu$}_{x}-A_{st}\mbox{\boldmath $\mu$}_{t}$ and $\mbox{\boldmath $\mu$}_{t{\cdot}{\rm pa}(t)}=(I_{n_t,n_t}-A_{tt})\mbox{\boldmath $\mu$}_{t}$
from Equation (\ref{a3}), we obtain
\begin{eqnarray*}
\lefteqn{(I_{n_s,n_s}-A_{ss})(\mbox{\boldmath $S$}-\mbox{\boldmath $\mu$}_{s})=A_{sx}(\mbox{\boldmath $X$}-\mbox{\boldmath $\mu$}_{x})+A_{st}(\mbox{\boldmath $T$}-\mbox{\boldmath $\mu$}_{t})+\mbox{\boldmath $\epsilon$}_{s.r_{h}}} \\
&=&A_{sx}(\mbox{\boldmath $x$}+C_{xs}\mbox{\boldmath $S$}+C_{xt}\mbox{\boldmath $T$}+\mbox{\boldmath $\epsilon$}^{\ast}_{x}-\mbox{\boldmath $\mu$}_{x})+A_{st}(\mbox{\boldmath $T$}-\mbox{\boldmath $\mu$}_{t})+\mbox{\boldmath $\epsilon$}_{s.r_{h}}\\
&=&A_{sx}(\mbox{\boldmath $x$}+C_{xs}\mbox{\boldmath $S$}-\mbox{\boldmath $\mu$}_{x})+(A_{st}+A_{sx}C_{xt})\mbox{\boldmath $T$}-A_{st}\mbox{\boldmath $\mu$}_{t}+A_{sx}\mbox{\boldmath $\epsilon$}^{\ast}_{x}+\mbox{\boldmath $\epsilon$}_{s.r_{h}}\\
&=&A_{sx}(\mbox{\boldmath $x$}+C_{xs}\mbox{\boldmath $S$}-\mbox{\boldmath $\mu$}_{x})+(A_{st}+A_{sx}C_{xt})(I_{n_t,n_t}-A_{tt})^{-1}(\mbox{\boldmath $\mu$}_{t.pa(t)}+\mbox{\boldmath $\epsilon$}_{t.{r_h}})-A_{st}\mbox{\boldmath $\mu$}_{t}\\
&&+A_{sx}\mbox{\boldmath $\epsilon$}^{\ast}_{x}+\mbox{\boldmath $\epsilon$}_{s.r_{h}}\\
&=&A_{sx}(\mbox{\boldmath $x$}+C_{xs}\mbox{\boldmath $S$}-\mbox{\boldmath $\mu$}_{x})+(A_{st}+A_{sx}C_{xt})(\mbox{\boldmath $\mu$}_{t}+(I_{n_t,n_t}-A_{tt})^{-1}\mbox{\boldmath $\epsilon$}_{t.{r_h}})-A_{st}\mbox{\boldmath $\mu$}_{t}\\
&&+A_{sx}\mbox{\boldmath $\epsilon$}^{\ast}_{x}+\mbox{\boldmath $\epsilon$}_{s.r_{h}}\\
&=&A_{sx}(\mbox{\boldmath $x$}+C_{xs}\mbox{\boldmath $S$}-\mbox{\boldmath $\mu$}_{x})+A_{sx}C_{xt}\mbox{\boldmath $\mu$}_{t}+(A_{st}+A_{sx}C_{xt})(I_{n_t,n_t}-A_{tt})^{-1}\mbox{\boldmath $\epsilon$}_{t.r_h}\\
&&+A_{sx}\mbox{\boldmath $\epsilon$}^{\ast}_{x}+\mbox{\boldmath $\epsilon$}_{s.r_{h}},
\end{eqnarray*}
that is, 
\begin{eqnarray*}
\lefteqn{(I_{n_s,n_s}-A_{ss}-A_{sx}C_{xs})\mbox{\boldmath $S$}=(I_{n_s,n_s}-A_{ss}) \mbox{\boldmath $\mu$}_{s}+A_{sx}(\mbox{\boldmath $x$}-\mbox{\boldmath $\mu$}_{x})+A_{sx}C_{xt}\mbox{\boldmath $\mu$}_{t}} \\
&&+(A_{st}+A_{sx}C_{xt})(I_{n_t,n_t}-A_{tt})^{-1}\mbox{\boldmath $\epsilon$}_{t.r_h}+A_{sx}\mbox{\boldmath $\epsilon$}^{\ast}_{x}+\mbox{\boldmath $\epsilon$}_{s.r_{h}}.
\end{eqnarray*}
Thus, letting $\mbox{\boldmath $\tau$}_{sx}=(I_{n_s,n_s}-A_{ss})^{-1}A_{sx}$, since we have
\begin{eqnarray*}
\lefteqn{(I_{n_s,n_s}-\mbox{\boldmath $\tau$}_{sx}C_{xs})\mbox{\boldmath $S$}=\mbox{\boldmath $\tau$}_{sx}(\mbox{\boldmath $x$}-\mbox{\boldmath $\mu$}_{x})+\mbox{\boldmath $\mu$}_{s}+\mbox{\boldmath $\tau$}_{sx}C_{xt}\mbox{\boldmath $\mu$}_{t}+(I_{n_s,n_s}-A_{ss})^{-1}}
 \nonumber \\
&&\times (A_{st}+A_{sx}C_{xt})(I_{n_t,n_t}-A_{tt})^{-1}\mbox{\boldmath $\epsilon$}_{t.r_h}+\mbox{\boldmath $\tau$}_{sx}\mbox{\boldmath $\epsilon$}^{\ast}_{x}+(I_{n_s,n_s}-A_{ss})^{-1}\mbox{\boldmath $\epsilon$}_{s.r_{h}},
\end{eqnarray*}
the modified SEM for $\mbox{\boldmath $S$}$ is given by
\begin{eqnarray}
\mbox{\boldmath $S$}&=&
(I_{n_s,n_s}-\mbox{\boldmath $\tau$}_{sx}C_{xs})^{-1}\left\{\mbox{\boldmath $\tau$}_{sx}(\mbox{\boldmath $x$}-\mbox{\boldmath $\mu$}_{x})+\mbox{\boldmath $\mu$}_{s}+\mbox{\boldmath $\tau$}_{sx}C_{xt}\mbox{\boldmath $\mu$}_{t}+(I_{n_s,n_s}-A_{ss})^{-1}\right.
 \nonumber \\
&&\times (A_{st}+A_{sx}C_{xt})(I_{n_t,n_t}-A_{tt})^{-1}\mbox{\boldmath $\epsilon$}_{t.r_h}+\mbox{\boldmath $\tau$}_{sx}\mbox{\boldmath $\epsilon$}^{\ast}_{x}+\left. (I_{n_s,n_s}-A_{ss})^{-1}\mbox{\boldmath $\epsilon$}_{s.r_{h}}\right\}.
\label{a10}
\end{eqnarray}

\subsection{Step 3: Prediction}

Finally, noting that 
\begin{eqnarray*}
\mbox{\boldmath $\mu$}_{\epsilon_{t}.r_h}&=&(I_{n_t,n_t}-A_{tt})(\mbox{\boldmath $\mu$}_{{t}.r_h}-\mbox{\boldmath $\mu$}_{t}),\,\,\,\, 
\mbox{\boldmath $\mu$}_{\epsilon^{*}_x}=\mbox{\boldmath $0$}_{n_x,1}, \\
\mbox{\boldmath $\mu$}_{\epsilon_{s}.r_h}&=&(I_{n_s,n_s}-A_{ss})(\mbox{\boldmath $\mu$}_{{s}.r_h}-\mbox{\boldmath $\mu$}_{s})-A_{sx}(\mbox{\boldmath $\mu$}_{{x}.r_h}-\mbox{\boldmath $\mu$}_{x})-A_{st}(\mbox{\boldmath $\mu$}_{{t}.r_h}-\mbox{\boldmath $\mu$}_{t})
\end{eqnarray*}
from Equation (\ref{313}), we have
\begin{eqnarray}
\lefteqn{E(\mbox{\boldmath $S$}|{\rm do}(\mbox{\boldmath $X$}=h(\mbox{\boldmath $F$},\mbox{\boldmath $W$},\mbox{\boldmath $\epsilon$}^{\ast}_{x})),\mbox{\boldmath $H$}\in R_{h})
=(I_{n_s,n_s}-\mbox{\boldmath $\tau$}_{sx}C_{xs})^{-1}\left\{\mbox{\boldmath $\tau$}_{sx}(\mbox{\boldmath $x$}-\mbox{\boldmath $\mu$}_{x})+\mbox{\boldmath $\mu$}_{s}\right. } \nonumber \\
&&+\mbox{\boldmath $\tau$}_{sx}C_{xt}\mbox{\boldmath $\mu$}_{t}+(I_{n_s,n_s}-A_{ss})^{-1}(A_{st}+A_{sx}C_{xt})(I_{n_t,n_t}-A_{tt})^{-1}\mbox{\boldmath $\mu$}_{\epsilon_{t}.r_h} \nonumber\\
&&\left. +(I_{n_s,n_s}-A_{ss})^{-1}\mbox{\boldmath $\mu$}_{\epsilon_s.r_{h}}\right\} \nonumber \\
&=&(I_{n_s,n_s}-\mbox{\boldmath $\tau$}_{sx}C_{xs})^{-1}\left\{\mbox{\boldmath $\tau$}_{sx}(\mbox{\boldmath $x$}-\mbox{\boldmath $\mu$}_{x})+\mbox{\boldmath $\mu$}_{s}+\mbox{\boldmath $\tau$}_{sx}C_{xt}\mbox{\boldmath $\mu$}_{t}\right.  \nonumber \\
&&+(I_{n_s,n_s}-A_{ss})^{-1}(A_{st}+A_{sx}C_{xt})(I_{n_t,n_t}-A_{tt})^{-1}(I_{n_t,n_t}-A_{tt})(\mbox{\boldmath $\mu$}_{{t}.r_h}-\mbox{\boldmath $\mu$}_{t}) \nonumber\\
&&\left. +(I_{n_s,n_s}-A_{ss})^{-1}((I_{n_s,n_s}-A_{ss})(\mbox{\boldmath $\mu$}_{{s}.r_h}-\mbox{\boldmath $\mu$}_{s})-A_{sx}(\mbox{\boldmath $\mu$}_{{x}.r_h}-\mbox{\boldmath $\mu$}_{x})-A_{st}(\mbox{\boldmath $\mu$}_{{t}.r_h}-\mbox{\boldmath $\mu$}_{t}))\right\} \nonumber \\
&=&(I_{n_s,n_s}-\mbox{\boldmath $\tau$}_{sx}C_{xs})^{-1}\left\{
\mbox{\boldmath $\tau$}_{sx}\mbox{\boldmath $x$}+(I_{n_s,n_s},-\mbox{\boldmath $\tau$}_{sx},\mbox{\boldmath $\tau$}_{sx}C_{xt})\mbox{\boldmath $\mu$}_{v.r_h}\right\}\label{a11}
\end{eqnarray}
and
\begin{eqnarray}
\lefteqn{\mbox{var}(\mbox{\boldmath $S$}|{\rm do}(\mbox{\boldmath $X$}=h(\mbox{\boldmath $F$},\mbox{\boldmath $W$},\mbox{\boldmath $\epsilon$}^{\ast}_{x})),\mbox{\boldmath $H$}\in R_{h})}\nonumber \\
&=&(I_{n_s,n_s}-\mbox{\boldmath $\tau$}_{sx}C_{xs})^{-1}\mbox{\boldmath $\tau$}_{sx}\Sigma_{\epsilon^{\ast}_{x},\epsilon^{\ast}_{x}}\mbox{\boldmath $\tau$}'_{sx}(I_{n_s,n_s}-\mbox{\boldmath $\tau$}_{sx}C_{xs})'^{-1} \nonumber \\
&&+(I_{n_s,n_s}-\mbox{\boldmath $\tau$}_{sx}C_{xs})^{-1}
((I_{n_s,n_s}-A_{ss})^{-1}, (I_{n_s,n_s}-A_{ss})^{-1}(A_{st}+A_{sx}C_{xt})(I_{n_t,n_t}-A_{tt})^{-1}) \nonumber \\
&&\times \mbox{var}
\left(\begin{array}{c}
\mbox{\boldmath $\epsilon$}_{s.r_{h}}\\
\mbox{\boldmath $\epsilon$}_{t.r_h}
\end{array}
\right)
\left(
\begin{array}{c}
(I_{n_s,n_s}-A'_{ss})^{-1}\\
(I_{n_t,n_t}-A'_{tt})^{-1}(A'_{st}+C'_{xt}A'_{sx})(I_{n_s,n_s}-A'_{ss})^{-1}
\end{array}
\right)\nonumber \\
&&\times (I_{n_s,n_s}-\mbox{\boldmath $\tau$}_{sx}C_{xs})'^{-1} \nonumber \\
&=&(I_{n_s,n_s}-\mbox{\boldmath $\tau$}_{sx}C_{xs})^{-1}\mbox{\boldmath $\tau$}_{sx}\Sigma_{\epsilon^{\ast}_{x},\epsilon^{\ast}_{x}}\mbox{\boldmath $\tau$}'_{sx}(I_{n_s,n_s}-\mbox{\boldmath $\tau$}_{sx}C_{xs})'^{-1} \nonumber \\
&&+(I_{n_s,n_s}-\mbox{\boldmath $\tau$}_{sx}C_{xs})^{-1}
((I_{n_s,n_s}-A_{ss})^{-1}, (I_{n_s,n_s}-A_{ss})^{-1}(A_{st}+A_{sx}C_{xt})(I_{n_t,n_t}-A_{tt})^{-1}) \nonumber \\
&&\times
\left(\begin{array}{ccc}
I_{n_s,n_s}-A_{ss}& -A_{sx}&-A_{st}\\
0_{n_t,n_s}&0_{n_t,n_x}&I_{n_t,n_t}-A_{tt}
\end{array}
\right)\Sigma_{vv.r_{h}}
\left(\begin{array}{cc}
I_{n_s,n_s}-A'_{ss}&0_{n_t,n_s}\\
 -A'_{sx}&0_{n_t,n_x}\\
-A'_{st}&I_{n_t,n_t}-A'_{tt}
\end{array}
\right) \nonumber \\
&&\times \left(
\begin{array}{c}
(I_{n_s,n_s}-A'_{ss})^{-1}\\
(I_{n_t,n_t}-A'_{tt})^{-1}(A'_{st}+C'_{xt}A'_{sx})(I_{n_s,n_s}-A'_{ss})^{-1}
\end{array}
\right)(I_{n_s,n_s}-\mbox{\boldmath $\tau$}_{sx}C_{xs})'^{-1} \nonumber \\
&=&(I_{n_s,n_s}-\mbox{\boldmath $\tau$}_{sx}C_{xs})^{-1}\mbox{\boldmath $\tau$}_{sx}\Sigma_{\epsilon^{\ast}_{x},\epsilon^{\ast}_{x}}\mbox{\boldmath $\tau$}'_{sx}(I_{n_s,n_s}-\mbox{\boldmath $\tau$}_{sx}C_{xs})'^{-1} \nonumber \\
&&+(I_{n_s,n_s}-\mbox{\boldmath $\tau$}_{sx}C_{xs})^{-1}(I_{n_s,n_s},-\mbox{\boldmath $\tau$}_{sx},\mbox{\boldmath $\tau$}_{sx}C_{xt})\Sigma_{vv.r_h}(I_{n_s,n_s},-\mbox{\boldmath $\tau$}_{sx},\mbox{\boldmath $\tau$}_{sx}C_{xt})'\nonumber \\
&&\times (I_{n_s,n_s}-\mbox{\boldmath $\tau$}_{sx}C_{xs})'^{-1}
\label{a12}
\end{eqnarray}
from Equation (\ref{a10}). 

If both $C_{xs}$ and $C_{xt}$ are zero vectors and $R_{h}$ is given as a specific vector, i.e., $\mbox{\boldmath $H$}=\mbox{\boldmath $h$}$, these formulations provide an explicit expression of the counterfactual quantities that Balke and Pearl (1994a, 1995) called ``the mean vector and covariance matrix of variables in the counterfactual world under the plan do($\mbox{\boldmath $X$}=\mbox{\boldmath $x$}$)". 
It should be noted that the discussion by Balke and Pearl (1994a, 1995) is based on {point type knowledge} in the real world, whereas our results are applicable to both {point} and disjunctive knowledge in the real world. 
Thus, our results extend Balke and Pearl's counterfactual analysis in a framework of linear SEMs. 
Unlike Balke and Pearl (1995), neither the mean vector nor covariance matrix of random disturbances appears in our formulations, with the exception of $\Sigma_{\epsilon^{\ast}_{x},\epsilon^{\ast}_{x}}$, which is assumed to be a zero matrix by Balke and Pearl (1995) and Chan and Pearl (2015). 
In addition, because Balke and Pearl (1994) did not provide explicit expressions of these counterfactual quantities, it was necessary for practitioners to understand and perform the three steps to estimate the counterfactual quantities, which was not a trivial task. 
Our results address these difficulties, thereby reducing the computational effort expended by practitioners. 
Furthermore, it should be noted that Pearl (2009, pp.~389-391) also provided the mean of the response variable in the counterfactual world under the plan do($\mbox{\boldmath $X$}=\mbox{\boldmath $x$}$), which is valid for non-Gaussian random disturbances, including feedback loops. 
Our results, which are also valid for non-Gaussian random disturbances including feedback loops, can also be considered as an extension of Pearl's work (2009, pp.~389-391), that is, from an unconditional plan of a univariate treatment to a conditional plan of a set of treatments. 
Additionally, unlike Pearl's work (2009, pp.~389-391), which focuses on the mean vector of variables in the counterfactual world, we provide an explicit expression not only of the mean vector of variables but also of the covariance matrix of variables in the counterfactual world. 

\section{Disjunctive Plan, Stochastic Plan, and Compound Treatments}

Assuming that the cause-effect relationships between variables are represented by a recursive structural causal model and, for the sake of simplicity, the atomic plan $\mbox{do}(X=x+\epsilon^*_{x})$ is the focus of interest, we state the relationship between our control plan and the disjunctive plan discussed by Pearl (2017). 

First, note that our discussion is based on stochastic plans (Pearl, 2009, pp.~113-114) or the manipulation theorem (Spirtes et al., 2000, pp.~47-53), i.e., 
\begin{displaymath}
\mbox{pr}(y|\mbox{do}(X=x+\epsilon^*_{x}))=\sum_{\epsilon^*_{x}}\mbox{pr}(y|\mbox{do}(X=x+\epsilon^*_{x}),\epsilon^*_{x})\mbox{pr}^{*}(\epsilon^*_{x}),
\end{displaymath}
where $\epsilon^*_{x}$ determines the value of $X$ together with $x$. 
Thus, noting that $\epsilon^*_{x}$ in $\mbox{pr}(y|\mbox{do}(X=x+\epsilon^*_{x}),\epsilon^*_{x})$ can be considered as a constant value, if $\mbox{pr}(y|\mbox{do}(X=x),\epsilon^*_{x})$ is identifiable and $\mbox{pr}(\epsilon^{*}_x)$ is given, $\mbox{pr}(y|\mbox{do}(X=x+\epsilon^{\ast}_{x}))$ is also identifiable. 
Here, since $\epsilon^*_{x}$ is an exogenous variable which has an effect on $X$ only and is not affected by other variables, from {\it do} calculus, Rule 1 (insertion/deletion of observation) (Pearl, 2009), the identification problem of $\mbox{pr}(y|\mbox{do}(X=x+\epsilon^*_{x}),\epsilon^*_{x})$ is reduced to that of $\mbox{pr}(y|\mbox{do}(X=x))$. 

Second, letting $\mbox{pa}(X)$ be a set of the parents of $X$ and $\mbox{nd}(X)$ a set of non-descendants of $X$, note that the information of $\mbox{nd}(X)$ on $X$ can be summarized to that of $\mbox{pa}(X)$ on $X$, i.e., $\mbox{pr}(x|\mbox{nd}(X))=\mbox{pr}(x|\mbox{pa}(X))$ for any $X=x$. 
In addition, the worlds that are closest to the information of $\mbox{nd}(X)$ on $X$ have the same information of $\mbox{nd}(X)$ on $X$. 
Thus, regarding the disjunctive plan which is interpreted as the control plan that allows subjects to choose the value $x$ of $X$ from $x\in R_{x}$, denoted as $(X\in R_{x})$, the causal effect of $(X\in R_{x})$ on $Y$, $\mbox{pr}(y\backslash\hspace*{-1mm}\backslash (X \in R_{x}))$, is given by 
\begin{eqnarray*}
\lefteqn{\mbox{pr}(y\backslash\hspace*{-1mm}\backslash (X \in R_{x}))}\\
&=&\sum_{x\in R_x, v\backslash \{x,y\}}\frac{\mbox{pr}(x,y,\mbox{\boldmath $v$}\backslash \{x,y\})}{\displaystyle \sum_{x\in R_x}\mbox{pr}(x|\mbox{pa}(x))}
=\sum_{\scriptsize \mbox{pa}(x)}\mbox{pr}(y|x\in R_{x}, \mbox{pa}(x))\mbox{pr}(\mbox{pa}(x))\\
&=&\sum_{x\in R_x,{\scriptsize \mbox{pa}(x)}}\mbox{pr}(y|x,\mbox{pa}(x))\left(\frac{\mbox{pr}(x|\mbox{pa}(x))}{\displaystyle \sum_{x\in R_x}\mbox{pr}(x|\mbox{pa}(x))}\right)\mbox{pr}(\mbox{pa}(x))\\
&=&\sum_{x\in R_x, {\scriptsize \mbox{pa}(x)}}\mbox{pr}(y|x,\mbox{pa}(x))\mbox{pr}(x|\mbox{pa}(x)),x\in R_x)\mbox{pr}(\mbox{pa}(x))
\end{eqnarray*}
where 
\begin{displaymath}
\mbox{pr}(y|x\in R_{x}, \mbox{pa}(x))=\sum_{x\in R_x}\frac{\mbox{pr}(x, y, \mbox{pa}(x))}{\displaystyle \sum_{x\in R_x}\mbox{pr}(x\in R_{x}, \mbox{pa}(x))},
\,\,\,
\mbox{pr}(x|\mbox{pa}(x),x\in R_x)=\frac{\mbox{pr}(x|\mbox{pa}(x))}{\displaystyle \sum_{x\in R_x}\mbox{pr}(x|\mbox{pa}(x))}.
\end{displaymath}
In Pearl (2017), intuitively, this formula is also interpreted from the viewpoint of a stochastic plan with the stochastic policy that a subject observes the value of $\mbox{pa}(x)$ then chooses the action $\mbox{do}(x)$ from $R_{x}$ based on the probability $\mbox{pr}(x|\mbox{pa}(x),x \in R_{x})$ (Pearl, 2017). 

The main difference between the two control plans is that, in the former case, the assigned probability $\mbox{pr}^{*}(\epsilon^*_{x})$ is given based mainly on external knowledge (e.g., expert knowledge, pilot studies, or the controllability of $X$ in the actual situation), whereas $\mbox{pr}(x|\mbox{pa}(x),x \in R_{x})$ is generally evaluated within a main study in the latter case. 
In addition, the stochastic plan $\mbox{do}(X=x+\epsilon^*_{x})$ is the mathematical operation which removes the equation that nominally assigns values to variable $X$, and replaces it with a new equation, $X=x+\epsilon^*_{x}$ together with $\mbox{pr}(\epsilon^*_{x})$. 
Thus, noting that each equation in structural causal models represents a mathematical function that each input has a ``single" output (e.g., $X$ is assigned to a ``single" value if both values of $x$ and $\epsilon^*_x$ are given), such a plan can be discussed in the context of structural causal models. 
In contrast, by the definition,  such an equation does not allow for the ambiguity that $X$ can be taken a unspecified value in $R_x$; the disjunctive plan $(X\in R_{x})$ is not automatically formulated in the context of structural causal models, and thus ``closest worlds" semantics is used, together with the following two provisions: 
\vspace*{2mm}

{\noindent}\hspace*{\fill}\begin{minipage}{12cm}
\baselineskip 8mm

{\noindent}Provision 1: worlds with equal histories should be considered equally similar to any given world.

{\noindent}Provision 2: equally-similar worlds should receive mass in proportion to their prior probabilities.
\end{minipage}\hspace*{\fill}
\vspace*{4mm}

{\noindent}For the details on the disjunctive plan, refer to Pearl (2017). 
Furthermore, in the case of the disjunctive plan, because the constraint $X\in R_x$ is imposed on the treatment $X$ itself, it may not clarify which of the direct causes, error terms, or both causes such a constraint, possibly making it difficult to handle mathematical operations. 
This difficulty leads to that, for example,  $X$ d-separates $Y$ from $\mbox{pa}(X)$ but $\mbox{pr}(y|x\in R_x,\mbox{pa}(x))=\mbox{pr}(y|x\in R_x)$ may not hold in general. 

As a similar problem to ours, we would also like to state the relationship between our control plan and ``compound treatments" or ``multiple versions of treatment" (Hernan and VanderWeele, 2011; Petersen, 2011; VanderWeele and Hernan, 2013). 
Letting $D_{x^*}=\{x+\epsilon^*_x|\epsilon^*_x\in D_{\epsilon^*_x}\}$ be a set of versions for treatment $X=x$ and $X^*$ be a variable taking its value from $D_{x^*}$, the causal effect of treatment $X=x$ and version $x^*\in D_{x^*}$ is given by $\mbox{pr}(y|\mbox{do}(X=x),\mbox{do}(X^*=x^*))$, which is often discussed in the context of the causal effect of joint interventions. 
However, in contrast to VanderWeele and Hernan's statement ``once the version is known the treatment is also known" (VanderWeele and Hernan, 2013), 
our control plan allows us to overlap between the versions for any distinct treatments $x$ and $x' $ (i.e., $D^{x^*}\cap D^{x'^*}\neq \phi$ for $x,x'\in D_{x}$); therefore, the treatment may be given regardless of the version. 
On the contrary, if the data generating process from $X$ to $X^{*}$ is given and overlap between the versions for distinct treatments is allowed, based on the idea of transportability introduced by Hernan and VanderWeele (2011), some of our problems may be covered in their framework. For the details on the transportability, refer to Bareinboim and Pearl (2012) and Pearl and Bareinboim (2011,2014). 

\section{Mean and Variance of the Response Variable in the Counterfactual World}

From Equations (\ref{a11}) and (\ref{a12}), the following theorem is obtained: 

\noindent{\bf Theorem 1}
{\it 
In a stable linear SEM, ${\rm var}(Y|{\rm do}(\mbox{\boldmath $X$}=h(Y,\mbox{\boldmath $W$},\mbox{\boldmath $\epsilon$}^{\ast}_{x})),\mbox{\boldmath $H$}\in R_{h})$ is minimized if $\mbox{\boldmath $b$}$ satisfies
\begin{equation}
\mbox{\boldmath $\tau$}_{yx}(I_{n_x,n_x}-\mbox{\boldmath $a$}\mbox{\boldmath $\tau$}_{fx})^{-1}\mbox{\boldmath $a$}(\mbox{\boldmath $\tau$}_{fx}(\mbox{\boldmath $b$}-B_{xw.r_h})+B_{fw.r_h})+\mbox{\boldmath $\tau$}_{yx}(\mbox{\boldmath $b$}-B_{xw.r_h})+B_{yw.r_h}=\mbox{\boldmath $0$}_{1,n_w},
\label{5}
\end{equation}
for $\mbox{\boldmath $a$}$ such that the maximum value of the absolute value of all eigenvalues of matrix $\mbox{\boldmath $\tau$}_{fx}\mbox{\boldmath $a$}$ is less than one. 
Here, $\mbox{\boldmath $\tau$}_{yx}$ is given by the row corresponding to $Y$ in $\mbox{\boldmath $\tau$}_{sx}=(I_{n_s,n_s}-A_{ss})^{-1}A_{sx}$, and $\mbox{\boldmath $\tau$}_{fx}$ is given by the first $n_{f}$ rows of $\mbox{\boldmath $\tau$}_{sx}$. 
In addition, $B_{sx.r_h}=\Sigma_{sx.r_h}\Sigma^{-1}_{xx.r_h}$, $B_{sw.r_h}=\Sigma_{sw.r_h}\Sigma^{-1}_{ww.r_h}$, and $B_{xw.r_h}=\Sigma_{xw.r_h}\Sigma^{-1}_{ww.r_h}$. 
Let $\mbox{\boldmath $b$}^{\ast}$ be the $\mbox{\boldmath $b$}$ satisfying Equation (\ref{5}) with the corresponding control plan ${\rm do}(\mbox{\boldmath $X$}=g(\mbox{\boldmath $F$},\mbox{\boldmath $W$},\mbox{\boldmath $\epsilon$}^{\ast}_{x}))$; then, we have
\begin{eqnarray*}
\lefteqn{E(Y|{\rm do}(\mbox{\boldmath $X$}=g(\mbox{\boldmath $F$},\mbox{\boldmath $W$},\mbox{\boldmath $\epsilon$}^{\ast}_{x})),\mbox{\boldmath $H$}\in R_{h})=\mu_{y.r_h}+\mbox{\boldmath $\tau$}_{yx}(\mbox{\boldmath $x$}-\mbox{\boldmath $\mu$}_{x.r_h}+\mbox{\boldmath $b$}^{\ast}\mbox{\boldmath $\mu$}_{w.r_h})}\nonumber \\
&&+\mbox{\boldmath $\tau$}_{yx}(I_{n_x,n_x}-\mbox{\boldmath $a$}\mbox{\boldmath $\tau$}_{fx})^{-1}
\mbox{\boldmath $a$}(\mbox{\boldmath $\mu$}_{f.r_h}+\mbox{\boldmath $\tau$}_{fx}(\mbox{\boldmath $x$}-\mbox{\boldmath $\mu$}_{x.r_h}+\mbox{\boldmath $b$}^{\ast}\mbox{\boldmath $\mu$}_{w.r_h}))
\label{25}
\end{eqnarray*}
and 
\begin{eqnarray*}
\lefteqn{\mbox{var}(Y|{\rm do}(\mbox{\boldmath $X$}=g(\mbox{\boldmath $F$},\mbox{\boldmath $W$},\mbox{\boldmath $\epsilon$}^{\ast}_{x})),\mbox{\boldmath $H$}\in R_{h})=\sigma^{\ast}_{yy}-\Sigma^{\ast}_{yf}\Sigma^{\ast -1}_{ff}\Sigma^{\ast}_{fy}} \label{a1}\\
&&+\left\{\mbox{\boldmath $\tau$}_{yx}(I_{n_x,n_x}-\mbox{\boldmath $a$}\mbox{\boldmath $\tau$}_{fx})^{-1}\mbox{\boldmath $a$}+\Sigma^{\ast}_{yf}\Sigma^{\ast -1}_{ff}\right\}\Sigma^{\ast }_{ff}\left\{\mbox{\boldmath $\tau$}_{yx}(I_{n_x,n_x}-\mbox{\boldmath $a$}\mbox{\boldmath $\tau$}_{fx})^{-1}\mbox{\boldmath $a$}+\Sigma^{\ast}_{yf}\Sigma^{\ast -1}_{ff}\right\}',\nonumber 
\end{eqnarray*}
where 
\begin{eqnarray*}
\Sigma^{\ast}_{ss}
&=&\left(\begin{array}{cc}
\Sigma^{\ast}_{ff}&\Sigma^{\ast}_{fu}\\
\Sigma^{\ast}_{uf}&\Sigma^{\ast}_{uu}
\end{array}
\right)=\Sigma_{ss.r_h}+\mbox{\boldmath $\tau$}_{sx}\Sigma_{\epsilon^{\ast}_{x} \epsilon^{\ast}_{x}}\mbox{\boldmath $\tau$}'_{sx}-B_{sx.r_h}\Sigma_{xx.r_h}B'_{sx.r_h}\\
&&+(\mbox{\boldmath $\tau$}_{sx}-B_{sx.r_h})\Sigma_{xx.r_h}(\mbox{\boldmath $\tau$}_{sx}-B_{sx.r_h})'\\
&&-(B_{sw.r_h}-\mbox{\boldmath $\tau$}_{sx}B_{xw.r_h})\Sigma_{ww}(B_{sw.r_h}-\mbox{\boldmath $\tau$}_{sx}B_{xw.r_h})'
\end{eqnarray*}
and $\sigma^{\ast}_{yy}$ is an element of $\Sigma^{\ast}_{ss}$ corresponding to $Y$. 
}
\hspace*{\fill}

The proof is given in Appendix 1. 
Here, do($X=g(\mbox{\boldmath $F$},\mbox{\boldmath $W$},\mbox{\boldmath $\epsilon$}^{\ast}_{x})$) in Theorem 1 is called an optimal plan of $\mbox{\boldmath $X$}$ for a given $\mbox{\boldmath $a$}$, because we can both adjust the value of the response variable to the target value and minimize its variance under the condition that $\mbox{\boldmath $F$}\cup\mbox{\boldmath $W$}$ is used for control. 
Additionally, to configure an optimal plan of $\mbox{\boldmath $X$}$ for a given $\mbox{\boldmath $a$}$, $\mbox{\boldmath $b$}$ can be determined by solving Equation (12) for a given $\mbox{\boldmath $a$}$. 
However, the choice of $\mbox{\boldmath $a$}$ is dependent on both the stability condition and the desired variance of the response variable. 

From Theorem 1, if we wish to configure a control plan that has the effect of bringing $Y$ close to target value $y_{0}$, the mean of $Y$ may be set to this value by selecting $\mbox{\boldmath $x$}$ satisfying 
\begin{eqnarray*}
\lefteqn{\mu_{y.r_h}+\mbox{\boldmath $\tau$}_{yx}(\mbox{\boldmath $x$}-\mbox{\boldmath $\mu$}_{x.r_h}+\mbox{\boldmath $b$}^{\ast}\mbox{\boldmath $\mu$}_{w.r_h})}\\
&&+\mbox{\boldmath $\tau$}_{yx}(I_{n_x,n_x}-\mbox{\boldmath $a$}\mbox{\boldmath $\tau$}_{fx})^{-1}
\mbox{\boldmath $a$}(\mbox{\boldmath $\mu$}_{f.r_h}+\mbox{\boldmath $\tau$}_{fx}(\mbox{\boldmath $x$}-\mbox{\boldmath $\mu$}_{x.r_h}+\mbox{\boldmath $b$}^{\ast}\mbox{\boldmath $\mu$}_{w.r_h}))=y_0
\end{eqnarray*}
under the assumptions in Theorem 1. 

In addition, Kuroki (2012) showed the mean vector and covariance matrix of $\mbox{\boldmath $S$}$ when carrying out control plan (3) without prior knowledge (i.e., $R_{h}$ is empty). 
The results are obtained by replacing $\mbox{\boldmath $\mu$}_{v.r_h}$ by $\mbox{\boldmath $\mu$}_{v}$ in Equation (10), and $\Sigma_{vv.r_h}$ by $\Sigma_{vv}$ in Equation (11). 
In addition, if we can estimate the total effects $\mbox{\boldmath $\tau$}_{yx}$ and $\mbox{\boldmath $\tau$}_{fx}$, calculation of the mean vector and covariance matrix when carrying out the control plan of $\mbox{\boldmath $X$}$ given $\mbox{\boldmath $H$}\in R_h$, can be achieved through the conditional mean vector and conditional covariance matrix of the variables given $\mbox{\boldmath $H$}\in R_h$ together with $\mbox{\boldmath $\tau$}_{yx}$, $\mbox{\boldmath $\tau$}_{fx}$, and the original path diagram. 
This observation can help SEM researchers and practitioners reduce the computational effort of evaluating counterfactual quantities, and generalizes the results of Kuroki (2012) and Kuroki and Miyakawa (2003). 

Furthermore, it can be seen from Theorem 1 that it is sufficient to calculate the covariance matrix of $\mbox{\boldmath $F$}{\cup}\mbox{\boldmath $W$}{\cup}\mbox{\boldmath $X$}{\cup}\{Y\}$, and the total effects $\mbox{\boldmath $\tau$}_{yx}$ and $\mbox{\boldmath $\tau$}_{fx}$ to evaluate counterfactual quantities, whereas certain elements of $\mbox{\boldmath $U$}\cup\mbox{\boldmath $Z$}$ may be used to estimate $\mbox{\boldmath $\tau$}_{yx}$ and $\mbox{\boldmath $\tau$}_{fx}$; it is note necessary to concerned with the evaluation of all the path coefficients. 
Thus, when assuming a recursive SEM as the data generating process, the nonparametric identification conditions for total effects presented by Kuroki and Miyakawa (1999), Pearl (2009), Tian (2008), Tian and Pearl (2002), and Tian and Shpitser (2010) as well as the (linear) parametric identification conditions proposed by Bollen (1989), Brito (2003), Brito and Pearl (2002abc), Cai and Kuroki (2008), Drton et al. (2011), Foygel et al. (2012), Kuroki and Pearl (2014), Pearl (2009), and Tian (2004, 2005, 2007ab) can be used to evaluate the total effect. Here, ``a total effect is identifiable" means that the total effect can be determined uniquely from statistical parameters of observed variables, such as observed covariances or joint distributions. 

The following theorem is obtained directly from Theorem 1. 

\noindent{\bf Theorem 2}
{\it 
\begin{displaymath}
\mbox{cov}(Y,\mbox{\boldmath $W$}|{\rm do}(\mbox{\boldmath $X$}=g(\mbox{\boldmath $F$},\mbox{\boldmath $W$},\mbox{\boldmath $\epsilon$}^{\ast}_{x})),\mbox{\boldmath $H$}\in R_{h})=\mbox{\boldmath $0$}_{1,n_w}.
\end{displaymath}
}\hspace*{\fill}

The proof is given in Appendix 2. 
From Theorem 2, the optimal plan based on Theorem 1 can be interpreted as a control plan that removes the correlation between $\mbox{\boldmath $W$}$ and $Y$. 

\section{Conclusion}

Counterfactual reasoning is an important issue in many practical sciences, although the theory is less developed in a linear SEM framework. 
If cause--effect relationships between variables are represented by a linear (possibly non-recursive) SEM, to solve the problem of clarifying how the mean vector and covariance matrix would change if various treatments were controlled by the values of covariates, intermediate variables, and/or a response variable (counterfactually), and given that prior knowledge is available in the form of disjunctive knowledge of some variables (actually), based on the imperfect control plan, we extended the counterfactual framework provided by Balke and Pearl (1995) from an unconditional plan to a conditional one, and from {point type knowledge} to disjunctive knowledge. 
In addition, we clarified various properties of these formulas. 
Furtheremore, we discussed the relationships between the imperfect control plan and the disjunctive plan discussed by Pearl (2017). 
The results of this study can help SEM researchers and practitioners not only reduce the computational effort of evaluating counterfactual quantities, but also understand the causal mechanisms of how the distributional characteristics would change if certain treatments were controlled in a given subpopulation. 
The discussion in this paper should also promote the application and development of counterfactual reasoning theory. 

\section*{\bf Appendix}

\subsection*{\bf Appendix 1: Proof of Theorem 1}

First, from Equation (10) and noting that $C_{xs}=(\mbox{\boldmath $a$};\mbox{\boldmath $0$}_{n_x,n_u})$ and $C_{xt}=(\mbox{\boldmath $b$};\mbox{\boldmath $0$}_{n_x,n_z})$, $\mbox{\boldmath $a$}$ must satisfy the condition that the maximum values of the absolute values of all eigenvalues of matrix $\mbox{\boldmath $\tau$}_{sx}(\mbox{\boldmath $a$};\mbox{\boldmath $0$}_{n_x,n_u})$ are less than one to obtain a stable structural equation model.  
We can obtain
\begin{eqnarray*}
(I_{n_s,n_s}-\mbox{\boldmath $\tau$}_{sx}C_{xs})^{-1}&=&I_{n_s,n_s}+\mbox{\boldmath $\tau$}_{sx}(I_{n_x,n_x}-\mbox{\boldmath $a$}\mbox{\boldmath $\tau$}_{fx})^{-1}C_{xs}=\left(\begin{array}{cc}
D_{1}&0_{n_f,f_u}\\
D_{2}&I_{n_u}
\end{array}
\right),
\end{eqnarray*}
where 
\begin{eqnarray*}
D_{1}&=&I_{n_f,n_f}+\mbox{\boldmath $\tau$}_{fx}(I_{n_x,n_x}-\mbox{\boldmath $a$}\mbox{\boldmath $\tau$}_{fx})^{-1}\mbox{\boldmath $a$},\hspace*{3mm}
D_{2}=\mbox{\boldmath $\tau$}_{ux}(I_{n_x,n_x}-\mbox{\boldmath $a$}\mbox{\boldmath $\tau$}_{fx})^{-1}\mbox{\boldmath $a$}.
\end{eqnarray*}
In addition, $\mbox{\boldmath $\tau$}_{fx}$ and $\mbox{\boldmath $\tau$}_{ux}$ are the first $n_{f}$ rows and the next $n_{u}$ rows of $\mbox{\boldmath $\tau$}_{sx}$, respectively. 
Thus, we can derive
\begin{eqnarray*}
\lefteqn{E(\mbox{\boldmath $F$}|{\rm do}(\mbox{\boldmath $X$}=h(\mbox{\boldmath $F$},\mbox{\boldmath $W$},\mbox{\boldmath $\epsilon$}^{\ast}_{x})),\mbox{\boldmath $H$}\in R_{h})=D_{1}(\mbox{\boldmath $\mu$}_{f.r_h}+\mbox{\boldmath $\tau$}_{fx}(\mbox{\boldmath $x$}-\mbox{\boldmath $\mu$}_{x.r_h}+\mbox{\boldmath $b$}\mbox{\boldmath $\mu$}_{w.r_h}))}\\
\lefteqn{E(\mbox{\boldmath $U$}|{\rm do}(\mbox{\boldmath $X$}=h(\mbox{\boldmath $F$},\mbox{\boldmath $W$},\mbox{\boldmath $\epsilon$}^{\ast}_{x})),\mbox{\boldmath $H$}\in R_{h})}\\
&=&D_{2}(\mbox{\boldmath $\mu$}_{f.r_h}+\mbox{\boldmath $\tau$}_{fx}(\mbox{\boldmath $x$}-\mbox{\boldmath $\mu$}_{x.r_h}+\mbox{\boldmath $b$}\mbox{\boldmath $\mu$}_{w.r_h}))+\mbox{\boldmath $\mu$}_{u.r_h}+\mbox{\boldmath $\tau$}_{ux}(\mbox{\boldmath $x$}-\mbox{\boldmath $\mu$}_{x.r_h}+\mbox{\boldmath $b$}\mbox{\boldmath $\mu$}_{w.r_h}).
\end{eqnarray*}
Even if $Y$ is included in $\mbox{\boldmath $F$}$ or $\mbox{\boldmath $U$}$, the mean of $Y$ is given by
\begin{eqnarray*}
\lefteqn{E(Y|{\rm do}(\mbox{\boldmath $X$}=h(\mbox{\boldmath $F$},\mbox{\boldmath $W$},\mbox{\boldmath $\epsilon$}^{\ast}_{x})),\mbox{\boldmath $H$}\in R_{h})=\mbox{\boldmath $\tau$}_{yx}(\mbox{\boldmath $x$}-\mbox{\boldmath $\mu$}_{x.r_h}+\mbox{\boldmath $b$}\mbox{\boldmath $\mu$}_{w.r_h})+\mu_{y.r_h}}\\
&&+\mbox{\boldmath $\tau$}_{yx}(I_{n_x,n_x}-\mbox{\boldmath $a$}\mbox{\boldmath $\tau$}_{fx})^{-1}\mbox{\boldmath $a$}(\mbox{\boldmath $\mu$}_{f.r_h}+\mbox{\boldmath $\tau$}_{fx}(\mbox{\boldmath $x$}-\mbox{\boldmath $\mu$}_{x.r_h}+\mbox{\boldmath $b$}\mbox{\boldmath $\mu$}_{w.r_h})). 
\end{eqnarray*}

Next, from Equation (11), we obtain
\begin{eqnarray*}
\lefteqn{(I_{n_s,n_s}-\mbox{\boldmath $\tau$}_{sx}C_{xs}){\rm var}(\mbox{\boldmath $S$}|{\rm do}(\mbox{\boldmath $X$}=h(\mbox{\boldmath $F$},\mbox{\boldmath $W$},\mbox{\boldmath $\epsilon$}^{\ast}_{x})),\mbox{\boldmath $H$}\in R_{h})(I_{n_s,n_s}-\mbox{\boldmath $\tau$}_{sx}C_{xs})'} \nonumber \\
&=&\Sigma_{ss.r_h}-B_{sx.r_h}\Sigma_{xx.r_h}B_{sx.r_h}+(\mbox{\boldmath $\tau$}_{sx}-B_{sx.r_h})\Sigma_{xx.r_h}(\mbox{\boldmath $\tau$}_{sx}-B_{sx.r_h})'+\mbox{\boldmath $\tau$}_{sx}\Sigma_{\epsilon^{\ast}_{x} \epsilon^{\ast}_{x}}\mbox{\boldmath $\tau$}'_{sx}\\
&&+(\mbox{\boldmath $\tau$}_{sx}\mbox{\boldmath $b$}+B_{sw.r_h}-\mbox{\boldmath $\tau$}_{sx}B_{xw.r_h})\Sigma_{ww.r_h}(\mbox{\boldmath $\tau$}_{sx}\mbox{\boldmath $b$}+B_{sw.r_h}-\mbox{\boldmath $\tau$}_{sx}B_{xw.r_h})'\\
&&-(B_{sw.r_h}-\mbox{\boldmath $\tau$}_{sx}B_{xw.r_h})\Sigma_{ww.r_h}(B_{sw.r_h}-\mbox{\boldmath $\tau$}_{sx}B_{xw.r_h})'. 
\end{eqnarray*}
Here, letting
\begin{eqnarray*}
\Sigma^{\ast}_{ss}&=&\Sigma_{ss.r_h}+\mbox{\boldmath $\tau$}_{sx}\Sigma_{\epsilon^{\ast}_{x} \epsilon^{\ast}_{x}}\mbox{\boldmath $\tau$}'_{sx}-B_{sx.r_h}\Sigma_{xx.r_h}B_{sx.r_h}+(\mbox{\boldmath $\tau$}_{sx}-B_{sx.r_h})\Sigma_{xx.r_h}(\mbox{\boldmath $\tau$}_{sx}-B_{sx.r_h})'\\
&&-(B_{sw.r_h}-\mbox{\boldmath $\tau$}_{sx}B_{xw.r_h})\Sigma_{ww.r_h}(B_{sw.r_h}-\mbox{\boldmath $\tau$}_{sx}B_{xw.r_h})', 
\end{eqnarray*}
we have
\begin{eqnarray*}
\lefteqn{{\rm var}(\mbox{\boldmath $F$}|{\rm do}(\mbox{\boldmath $X$}=h(\mbox{\boldmath $F$},\mbox{\boldmath $W$},\mbox{\boldmath $\epsilon$}^{\ast}_{x})),\mbox{\boldmath $H$}\in R_{h})}\\
&=&D_{1}\left\{\frac{}{}\Sigma^{\ast}_{ff}+(\mbox{\boldmath $\tau$}_{fx}\mbox{\boldmath $b$}+B_{fw.r_h}-\mbox{\boldmath $\tau$}_{fx}B_{xw.r_h})\Sigma_{ww.r_h}(\mbox{\boldmath $\tau$}_{fx}\mbox{\boldmath $b$}+B_{fw.r_h}-\mbox{\boldmath $\tau$}_{fx}B_{xw.r_h})'\frac{}{}\right\}D'_{1},\\
\lefteqn{{\rm var}(\mbox{\boldmath $U$}|{\rm do}(\mbox{\boldmath $X$}=h(\mbox{\boldmath $F$},\mbox{\boldmath $W$},\mbox{\boldmath $\epsilon$}^{\ast}_{x})),\mbox{\boldmath $H$}\in R_{h})}\\
&=&D_{2}\left\{\frac{}{}\Sigma^{\ast}_{ff}+(\mbox{\boldmath $\tau$}_{fx}\mbox{\boldmath $b$}+B_{fw.r_h}-\mbox{\boldmath $\tau$}_{sx}B_{xw.r_h})\Sigma_{ww.r_h}(\mbox{\boldmath $\tau$}_{fx}\mbox{\boldmath $b$}+B_{fw.r_h}-\mbox{\boldmath $\tau$}_{fx}B_{xw.r_h})'\frac{}{}\right\}D'_{2}\\
&&+\Sigma^{\ast}_{uu}+(\mbox{\boldmath $\tau$}_{ux}\mbox{\boldmath $b$}+B_{uw.r_h}-\mbox{\boldmath $\tau$}_{sx}B_{xw.r_h})\Sigma_{ww.r_h}(\mbox{\boldmath $\tau$}_{ux}\mbox{\boldmath $b$}+B_{uw.r_h}-\mbox{\boldmath $\tau$}_{ux}B_{xw.r_h})'\\
&&+\left\{\frac{}{}\Sigma^{\ast}_{uf}+(\mbox{\boldmath $\tau$}_{ux}\mbox{\boldmath $b$}+B_{uw.r_h}-\mbox{\boldmath $\tau$}_{ux}B_{xw.r_h})\Sigma_{ww.r_h}(\mbox{\boldmath $\tau$}_{fx}\mbox{\boldmath $b$}+B_{fw.r_h}-\mbox{\boldmath $\tau$}_{fx}B_{xw.r_h})'\frac{}{}\right\}D_{2}\\
&&+D_{2}\left\{\frac{}{}\Sigma^{\ast}_{fu}+(\mbox{\boldmath $\tau$}_{fx}\mbox{\boldmath $b$}+B_{fw.r_h}-\mbox{\boldmath $\tau$}_{fx}B_{xw.r_h})\Sigma_{ww.r_h}(\mbox{\boldmath $\tau$}_{ux}\mbox{\boldmath $b$}+B_{uw.r_h}-\mbox{\boldmath $\tau$}_{ux}B_{xw.r_h})'\frac{}{}\right\}\\
&=&\Sigma^{\ast}_{uu}-\Sigma^{\ast}_{uf}\Sigma^{\ast -1}_{ff}\Sigma^{\ast}_{fu}+(D_{2}+\Sigma^{\ast}_{uf}\Sigma^{\ast -1}_{ff})\Sigma^{\ast}_{ff}(D_{2}+\Sigma^{\ast}_{uf}\Sigma^{\ast -1}_{ff})'\\
&&+\left\{\frac{}{}D_{2}(\mbox{\boldmath $\tau$}_{fx}\mbox{\boldmath $b$}+B_{fw.r_h}-\mbox{\boldmath $\tau$}_{fx}B_{xw.r_h})+\mbox{\boldmath $\tau$}_{ux}\mbox{\boldmath $b$}+B_{uw.r_h}-\mbox{\boldmath $\tau$}_{ux}B_{xw.r_h}\frac{}{}\right\}\Sigma_{ww.r_h}\\
&&{\times}\left\{\frac{}{}D_{2}(\mbox{\boldmath $\tau$}_{fx}\mbox{\boldmath $b$}+B_{fw.r_h}-\mbox{\boldmath $\tau$}_{fx}B_{xw.r_h})+\mbox{\boldmath $\tau$}_{ux}\mbox{\boldmath $b$}+B_{uw.r_h}-\mbox{\boldmath $\tau$}_{ux}B_{xw.r_h})\frac{}{}\right\}.
\end{eqnarray*}
Here, we assume that $Y$ is the first component of $\mbox{\boldmath $F$}$. 
Then, regarding var$(\mbox{\boldmath $F$}|{\rm do}(\mbox{\boldmath $X$}=h(\mbox{\boldmath $F$},\mbox{\boldmath $W$},\mbox{\boldmath $\epsilon$}^{\ast}_{x})),\mbox{\boldmath $H$}\in R_{h})$, by noting that the first row of $D_{1}$ is provided by $(1,0,\cdots,0)+\mbox{\boldmath $\tau$}_{yx}(I_{n_x,n_x}-\mbox{\boldmath $a$}\mbox{\boldmath $\tau$}_{fx})^{-1}\mbox{\boldmath $a$}$, to minimize the variance of $Y$ for a given $\mbox{\boldmath $a$}$ such that the maximum values of the absolute values of all eigenvalues of matrix $\mbox{\boldmath $\tau$}_{sx}(\mbox{\boldmath $a$};\mbox{\boldmath $0$}_{n_x,n_{u}})$ are less than one, we solve the following equation regarding $\mbox{\boldmath $b$}$: 
\begin{equation}
\mbox{\boldmath $\tau$}_{yx}(I_{n_x,n_x}-\mbox{\boldmath $a$}\mbox{\boldmath $\tau$}_{fx})^{-1}\mbox{\boldmath $a$}(\mbox{\boldmath $\tau$}_{fx}(\mbox{\boldmath $b$}-B_{xw.r_h})+B_{fw.r_h})+\mbox{\boldmath $\tau$}_{yx}(\mbox{\boldmath $b$}-B_{xw.r_h})+B_{yw.r_h}=\mbox{\boldmath $0$}_{1,n_w}.
\label{7}
\end{equation}
Thus, letting $\mbox{\boldmath $b$}^{\ast}$ be the $\mbox{\boldmath $b$}$ satisfying Equation (13), we can obtain
\begin{eqnarray*}
\lefteqn{\mbox{var}(Y|{\rm do}(\mbox{\boldmath $X$}=g(\mbox{\boldmath $F$},\mbox{\boldmath $W$},\mbox{\boldmath $\epsilon$}^{\ast}_{x})),\mbox{\boldmath $H$}\in R_{h})=\sigma^{\ast}_{yy}-\Sigma^{\ast}_{yf}\Sigma^{\ast -1}_{ff}\Sigma^{\ast}_{fy}} \nonumber\\
&&+\left\{\mbox{\boldmath $\tau$}_{yx}(I_{n_x,n_x}-\mbox{\boldmath $a$}\mbox{\boldmath $\tau$}_{fx})^{-1}\mbox{\boldmath $a$}+\Sigma^{\ast}_{yf}\Sigma^{\ast -1}_{ff}\right\}\Sigma^{\ast }_{ff}\left\{\mbox{\boldmath $\tau$}_{yx}(I_{n_x,n_x}-\mbox{\boldmath $a$}\mbox{\boldmath $\tau$}_{fx})^{-1}\mbox{\boldmath $a$}+\Sigma^{\ast}_{yf}\Sigma^{\ast -1}_{ff}\right\}'. \label{b1}
\end{eqnarray*}

Next, we assume that $Y$ is the first component of $\mbox{\boldmath $U$}$. 
Then, regarding var$(\mbox{\boldmath $U$}|{\rm do}(\mbox{\boldmath $X$}=h(\mbox{\boldmath $F$},\mbox{\boldmath $W$},\mbox{\boldmath $\epsilon$}^{\ast}_{x})),\mbox{\boldmath $H$}\in R_{h})$, by noting that the first row of $D_{2}$ is provided by $\mbox{\boldmath $\tau$}_{yx}(I_{n_x,n_x}-\mbox{\boldmath $a$}\mbox{\boldmath $\tau$}_{fx})^{-1}\mbox{\boldmath $a$}$,  to minimize the variance of $Y$ for a given $\mbox{\boldmath $a$}$ such that the maximum values of the absolute values of all eigenvalues of matrix $\mbox{\boldmath $\tau$}_{sx}(\mbox{\boldmath $a$};\mbox{\boldmath $0$}_{n_x,n_{u}})$ are less than one, we can solve Equation (\ref{7}) regarding $\mbox{\boldmath $b$}$, because the first row of $
D_{2}(\mbox{\boldmath $\tau$}_{fx}\mbox{\boldmath $b$}'+B_{fw.r_h}-\mbox{\boldmath $\tau$}_{fx}B_{xw.r_h})+\mbox{\boldmath $\tau$}_{ux}\mbox{\boldmath $b$}'+B_{uw.r_h}-\mbox{\boldmath $\tau$}_{ux}B_{xw.r_h}$ is equal to Equation (\ref{7}). 

\subsection*{\bf Appendix 2: Proof of Theorem 2}
We have
\begin{eqnarray*}
\lefteqn{(I_{n_s,n_s}-A_{ss}-A_{sx}C_{xs})\mbox{cov}(\mbox{\boldmath $S$}, \mbox{\boldmath $W$}|{\rm do}(\mbox{\boldmath $X$}=g(\mbox{\boldmath $F$},\mbox{\boldmath $W$},\mbox{\boldmath $\epsilon$}^{\ast}_{x})),\mbox{\boldmath $H$}\in R_{h})}\\
&=&A_{sx}\mbox{cov}(\mbox{\boldmath $\epsilon$}^{*}_{x}, \mbox{\boldmath $W$}|{\rm do}(\mbox{\boldmath $X$}=g(\mbox{\boldmath $F$},\mbox{\boldmath $W$},\mbox{\boldmath $\epsilon$}^{\ast}_{x})),\mbox{\boldmath $H$}\in R_{h}) \nonumber\\
&&+(A_{st}+A_{sx}C_{xt})\mbox{cov}(\mbox{\boldmath $T$}, \mbox{\boldmath $W$}|{\rm do}(\mbox{\boldmath $X$}=g(\mbox{\boldmath $F$},\mbox{\boldmath $W$},\mbox{\boldmath $\epsilon$}^{\ast}_{x})),\mbox{\boldmath $H$}\in R_{h})\\
&&+\mbox{cov}(\mbox{\boldmath $\epsilon$}_{s}, \mbox{\boldmath $W$}|{\rm do}(\mbox{\boldmath $X$}=g(\mbox{\boldmath $F$},\mbox{\boldmath $W$},\mbox{\boldmath $\epsilon$}^{\ast}_{x})),\mbox{\boldmath $H$}\in R_{h})\\
&=&(A_{st}+A_{sx}C_{xt})\mbox{cov}(\mbox{\boldmath $T$}, \mbox{\boldmath $W$}|{\rm do}(\mbox{\boldmath $X$}=g(\mbox{\boldmath $F$},\mbox{\boldmath $W$},\mbox{\boldmath $\epsilon$}^{\ast}_{x})),\mbox{\boldmath $H$}\in R_{h})\\
&&+\mbox{cov}(\mbox{\boldmath $\epsilon$}_{s}, \mbox{\boldmath $W$}|{\rm do}(\mbox{\boldmath $X$}=g(\mbox{\boldmath $F$},\mbox{\boldmath $W$},\mbox{\boldmath $\epsilon$}^{\ast}_{x})),\mbox{\boldmath $H$}\in R_{h})\\
&=&(A_{st}+A_{sx}C_{xt})\Sigma_{tw.r_h}+(I_{n_s,n_s}-A_{ss})\Sigma_{sw.r_h}-A_{sx}\Sigma_{xw.r_h}-A_{st}\Sigma_{tw.r_h}\\
&=&A_{sx}C_{xt}\Sigma_{tw.r_h}+(I_{n_s,n_s}-A_{ss})\Sigma_{sw.r_h}-A_{sx}\Sigma_{xw.r_h}.
\end{eqnarray*}
Thus, 
\begin{eqnarray*}
\lefteqn{\mbox{cov}(\mbox{\boldmath $S$}, \mbox{\boldmath $W$}|{\rm do}(\mbox{\boldmath $X$}=g(\mbox{\boldmath $F$},\mbox{\boldmath $W$},\mbox{\boldmath $\epsilon$}^{\ast}_{x})),\mbox{\boldmath $H$}\in R_{h})}\\
&=&(I_{n_s,n_s}-\mbox{\boldmath $\tau$}_{sx}C_{xs})^{-1}(\mbox{\boldmath $\tau$}_{sx}\mbox{\boldmath $b$}\Sigma_{ww.r_h}+\Sigma_{sw.r_h}-\mbox{\boldmath $\tau$}_{sx}\Sigma_{xw.r_h})\\
&=&(I_{n_s,n_s}-\mbox{\boldmath $\tau$}_{sx}C_{xs})^{-1}(\mbox{\boldmath $\tau$}_{sx}\mbox{\boldmath $b$}+B_{sw.r_h}-\mbox{\boldmath $\tau$}_{sx}B_{xw.r_h})\Sigma_{ww.r_h}. 
\end{eqnarray*}
Even if $Y$ is included in $\mbox{\boldmath $F$}$ or $\mbox{\boldmath $U$}$, we have
\begin{eqnarray*}
\lefteqn{\mbox{cov}(Y, \mbox{\boldmath $W$}|{\rm do}(\mbox{\boldmath $X$}=g(\mbox{\boldmath $F$},\mbox{\boldmath $W$},\mbox{\boldmath $\epsilon$}^{\ast}_{x})),\mbox{\boldmath $H$}\in R_{h})}\\
&=&(\mbox{\boldmath $\tau$}_{yx}(I_{n_x,n_x}-\mbox{\boldmath $a$}\mbox{\boldmath $\tau$}_{fx})^{-1}\mbox{\boldmath $a$}(\mbox{\boldmath $\tau$}_{fx}(\mbox{\boldmath $b$}^*-B_{xw.r_h})+B_{fw.r_h})+\mbox{\boldmath $\tau$}_{yx}(\mbox{\boldmath $b$}^*-B_{xw.r_h})+B_{yw.r_h})\Sigma_{ww.r_h}\\
&=&\mbox{\boldmath $0$}_{1,n_w}.
\end{eqnarray*}

\section*{Acknowledgements}

We would like to thank the chief editor and two anonymous reviewers whose comments significantly improved the presentation of this paper. 
This work was partially supported by the Ministry of Education, Culture, Sports, Science and Technology of Japan.

\section*{References}
\begin{list}{}{
\renewcommand{\makelabel}{\hfill}
\setlength{\itemindent}{-3mm}
\setlength{\topsep}{-1mm}
\setlength{\parsep}{-1mm}
\setlength{\leftmargin}{6mm}
}

\item Balke, A. and Pearl, J. (1994a). 
Probabilistic evaluation of counterfactual queries.  
{\it Proceedings of the 12th National Conference on Artificial Intelligence}, 230-237. 

\item Balke, A. and Pearl, J. (1994b). 
Counterfactual probabilities: Computational methods, bounds and identifications. 
{\it Proceeding of the 10th Conference on Uncertainty in Artificial Intelligence}, 11-18. 

\item Balke, A. and Pearl, J. (1995). 
Counterfactuals and policy analysis in structural models. 
{\it Proceeding of the 11th Conference on Uncertainty in Artificial Intelligence}, 46-54. 

\item Bareinboim, E. and Pearl, J. (2012). 
Transportability of Causal Effects: Completeness Results.
{\it Proceedings of the 26th AAAI Conference on Artificial Intelligence}, 698-704. 

\item Berry, W. D. (1984). 
{\it Nonrecursive causal models}. 
Sage Publications. 

\item 
Bentler, P. M. and Freeman, E. H. (1983). 
Tests for stability in linear structural equation systems. 
{\it Psychometrika}, {\bf 48}, 143--145. 

\item
Ben-Israel, A. and Greville, T. N. E. (1972). 
{\it Generalized Inverses: Theory and Applications}. Wiley.

\item Bollen, K. A. (1989). 
{\it Structural Equations with Latent Variables}. 
John Wiley \& Sons.

\item 
Bowden, R. J. , and Turkington, D. A. (1984). 
{\it Instrumental Variables}. 
Cambridge University Press. 

\item Brito, C. (2003). 
A new approach to the identification problem. 
{\it Advances in Artificial Intelligence: The 16th Brazilian Symposium on Artificial Intelligence}, 41-51. 

\item Brito, C. and Pearl, J. (2002a). 
Generalized instrumental variables. 
{\it Proceeding of the 18th Conference on Uncertainty in Artificial Intelligence}, 85-93. 

\item Brito, C. and Pearl, J. (2002b). 
A graphical criterion for the identification of causal effects in linear models. 
{\it Proceedings of the 18th National Conference on Artificial Intelligence}, 
533-538. 

\item Brito, C. and Pearl, J. (2002c). 
A new identification condition for recursive models with correlated errors. 
{\it Structural Equation Modeling: A Multidisciplinary Journal}, {\bf 9}, 459-474. 

\item 
Cai, Z. and Kuroki, M. (2005). 
Variance estimators for three ``probabilities of causation". 
{\it Risk Analysis}, {\bf 25}, 1611-1620.

\item 
Cai, Z. and Kuroki, M. (2008). 
On identifying total effects in the presence of latent variables and selection bias. 
{\it Proceeding of the 24th Conference on Uncertainty in Artificial Intelligence}, 62--69. 

\item 
Chen, B. and Pearl, J. (2014). Graphical tools for linear structural equation modeling. 
{\it Psychometrika}, Accepted. 

\item 
Dawid, A., Fienberg, S. and Faigman, D. (2014). 
Fitting science into legal contexts:
Assessing effects of causes or causes of effects? 
{\it Sociological Methods and Research}, {\bf 43}, 359-390.

\item 
Drton, M., Foygel, R., and Sullivant, S. (2011). 
Global identifiability of linear structural equation models. {\it Annals of Statistics}, {\bf 39}, 865-886.

\item 
Edwards, D. (2000). 
{\it Introduction to Graphical Modelling}. Springer.

\item Eusebia, P. (2008). 
A graphical method for assessing the identification of linear structural equation models. 
{\it Structural Equation Modeling: A Multidisciplinary Journal}, {\bf 15}, 403-412. 

\item 
Foygel, R., Draisma, J., and Drton, M. (2012). 
Half-trek criterion for generic identifiability of linear structural equation models. 
{\it Annals of Statistics}, {\bf 40}, 1682-1713.

\item Greenland, S. and Robins, J. M. (1988). 
Conceptual problems in the definition and interpretation of attributable fractions. 
{\it American Journal of Epidemiology}, \textbf{128}, 1185-1197. 

\item 
{Hernan, M. A., and VanderWeele, T. J. (2011). }
Compound Treatments and Transportability of Causal Inference. 
{\it Epidemiology}, {\bf 22}, 368-377.

\item 
Imbens, G. W. and Rubin, D. B. (2015). 
{\it Causal Inference in Statistics, Social, and Biomedical Sciences}. 
Cambridge University Press.

\item J$\ddot{\mbox{o}}$reskog, K. G. (1979). 
{\it Advances in Factor Analysis and Structural Equation Models}. 
Abt Books. 

\item Kuroki, M. (2012). 
Optimizing a control plan using causal diagram with an application to statistical process analysis. 
{\it Journal of Applied Statistics}, {\bf 39}, 673-694.

\item 
Kuroki, M. and Cai, Z. (2011). 
Statistical analysis of ``probabilities of causation" using covariate information. 
{\it Scandinavian Journal of Statistics}, {\bf 38}, 564-577.

\item Kuroki, M. and Miyakawa, M. (1999). 
Identifiability criteria for causal effects of joint interventions. 
{\it Journal of the Japan Statistical Society}, {\bf 29}, 105--117. 

\item 
{Kuroki, M. and Miyakawa, M. (2003). }
Covariate selection for estimating the causal effect of control plans by using causal diagrams. 
{\it Journal of the Royal Statistical Society: Series B (Statistical Methodology)}, {\bf 65}, 209-222.

\item 
Kuroki, M. and Pearl, J. (2014). 
Measurement bias and effect restoration in causal inference. 
{\it Biometrika}, {\bf 101}, 423-437.

\item 
Lauritzen, S. L. (1996). {\it Graphical Models}. 
Oxford University Press.

\item 
Morgan, S. L., and Winship, C. (2007). 
{\it Counterfactuals and Causal inference: Methods and Principles for Social Research}. 
Cambridge University Press.

\item Murphy, S.A. (2003). 
Optimal dynamic treatment regimes. 
{\it Journal of the Royal Statistical Society, Series B}, {\bf 65}, 331--366. 

\item Pearl, J. (1999). 
Probabilities of causation: Three counterfactual interpretations and their identification. 
{\it Synthese}, \textbf{121}, 93-149.

\item 
Pearl, J. (2009). 
{\it Causality: Models, Reasoning, and Inference}. 
The 2nd edition, Cambridge University Press. 

\item Pearl, J. (2017). 
Physical and Metaphysical Counterfactuals: Evaluating Disjunctive Action. 
UCLA Cognitive Systems Laboratory, Technical Report, R-359.

\item Pearl, J. and Bareinboim, E. (2011). 
Transportability across studies: A formal approach. 
{\it Proceedings of the 25th AAAI Conference on Artificial Intelligence}, 247-254. 

\item Pearl, J. and Bareinboim, E. (2014). 
External validity: From do-calculus to transportability across populations. 
{\it Statistical Science}, {\bf 29}, 579-595.

\item 
{Petersen, M. L. (2011). }
Compound treatments, transportability, and the structural causal model: The power and simplicity of causal graphs. 
{\it Epidemiology}, {\bf 22}, 378-381.

\item Robins, J. M. (2004). 
Should compensation schemes be based on the probability of causation or expected years of life lost? 
{\it Journal of Law and Policy}, \textbf{12}, 537-548.

\item Robins, J. M. and Greenland, S. (1989a). 
Estimability and estimation of excess and etiologic fractions. 
{\it Statistics in Medicine}, \textbf{8}, 845-859.

\item Robins, J. M. and Greenland, S. (1989b). 
The probability of causation under a stochastic model for individual risk. 
{\it Biometrics}, \textbf{45}, 1125-1138.

\item 
Rubin, D. (2006). 
{\it Matched Sampling for Casual Effects}. 
Cambridge University Press.

\item 
Spirtes, P., Glymour, C. N., and Scheines, R. (2000). 
{\it Causation, Prediction, and Search}. MIT press.

\item Tian, J. (2004). 
Identifying linear causal effects, 
{\it Proceedings the 19th National Conference on Artificial Intelligence
}, 104-110.

\item Tian, J. (2005). 
Identifying direct causal effects in linear models. 
{\it Proceedings of the 20th National Conference on Artificial Intelligence}, 346-352.

\item Tian, J. (2007a). 
On the identification of a class of linear models. 
{\it Proceedings of the 22nd National Conference on Artificial Intelligence}, 1284-1289.

\item Tian, J. (2007b). 
A criterion for parameter identification in structural equation models. 
{\it Proceedings of the 23rd Conference on Uncertainty in Artificial Intelligence}, 392-399.

\item Tian, J. (2008). 
Identifying dynamic sequential plans. 
{\it Proceedings of the 24rd Conference on Uncertainty in Artificial Intelligence}, 62--69.

\item Tian, J. and Pearl, J. (2000a). 
Probabilities of causation: Bounds and identification. 
{\it Annals of Mathematics and Artificial Intelligence}, {\bf 28}, 287-313. 

\item Tian, J. and Pearl, J. (2000b). 
Probabilities of causation: Bounds and identification. 
{\it Proceedings of the 16th Conference on Uncertainty in Artificial Intelligence}, 589-598. 

\item Tian, J. and Pearl, J. (2002). 
A general identification condition for causal effects. 
{\it Proceedings of the 18th National Conference on Artificial Intelligence}, 567--573. 

\item 
Tian, J. and Shpitser, I. (2010). 
On identifying causal effects. 
{\it Heuristics, Probability and Causality: A Tribute to Judea Pearl} (R. Dechter, H. Geffner and J. Halpern, eds.). College Publications, 415-444.

\item 
{VanderWeele, T. J. and Hernan, M. A. (2013). }
Causal inference under multiple versions of treatment. 
{\it Journal of causal inference}, {\bf 1}, 1-20.

\item 
Whittaker, J. (2009). {\it Graphical Models in Applied Multivariate Statistics}. Wiley.

\item Yamamoto, T. (2012). Understanding the past: Statistical analysis of causal attribution. 
{\it American Journal of Political Science}, {\bf 56}, 237--256.

\end{list}

\end{document}